\documentclass[pra,twocolumn,showpacs]{revtex4-1}
\usepackage{graphicx}
\usepackage{amsmath}
\usepackage{amsfonts}
\usepackage{amssymb}
\usepackage{epsfig}
\usepackage{color}
\usepackage{bm}
\usepackage[nice]{units}
\usepackage{hyperref}

\newcommand{\bra}[1]{\left\langle #1 \right|}
\newcommand{\ket}[1]{\left|#1\right\rangle}

\begin{document}

\title{Collective strong coupling between ion Coulomb crystals and an optical cavity field:\\ Theory
  and experiment}

\author{M. Albert}
\altaffiliation{Current address: Max-Planck-Institut für Quantenoptik,
  Hans-Kopfermann-Str. 1, 85748 Garching, Germany and
  Albert-Ludwigs-Universität Freiburg, Physikalisches Institut,
  Hermann-Herder-Str. 3, 79104 Freiburg, Germany}
\affiliation{QUANTOP, Danish National Research Foundation Center for Quantum Optics, Department of
Physics and Astronomy, University of Aarhus, DK-8000 \AA rhus C., Denmark}

\author{J. P. Marler}
\altaffiliation{Current address: Department of Physics and Astronomy,
  Northwestern University, 2145 Sheridan Road, Evanston IL 60208, USA} 
\affiliation{QUANTOP, Danish National Research Foundation Center for Quantum Optics, Department of
Physics and Astronomy, University of Aarhus, DK-8000 \AA rhus C., Denmark}

\author{P. F. Herskind}
\altaffiliation{Current address: Research Laboratory of Electronics,
  Massachusetts Institute of Technology, Cambridge, MA 02139, USA} 
\affiliation{QUANTOP, Danish National Research Foundation Center for Quantum Optics, Department of
Physics and Astronomy, University of Aarhus, DK-8000 \AA rhus C., Denmark}

\author{A. Dantan}
\affiliation{QUANTOP, Danish National Research Foundation Center for Quantum Optics, Department of
Physics and Astronomy, University of Aarhus, DK-8000 \AA rhus C., Denmark}

\author{M. Drewsen}\email{drewsen@phys.au.dk}
\affiliation{QUANTOP, Danish National Research Foundation Center for Quantum Optics, Department of
Physics and Astronomy, University of Aarhus, DK-8000 \AA rhus C., Denmark}

\begin{abstract}
A detailed description and theoretical analysis of experiments achieving coherent coupling
between an ion Coulomb crystal and an optical cavity field are presented. The various methods used
to measure the coherent coupling rate between large ion Coulomb crystals in a linear quadrupole
radiofrequency ion trap and a single-field mode of a moderately high-finesse
cavity are described in detail. Theoretical models based on a semiclassical 
approach are applied in assessment of the experimental results of [P.~F.~Herskind \textit{et al.},
Nature Phys. {\bf 5}, 494 (2009)] and of complementary new measurements. Generally, a very good
agreement between theory and experiments is obtained.
\end{abstract}


\date{\today}

\maketitle

\section{Introduction}
Cavity Quantum Electrodynamics (CQED) constitutes a fundamental platform for
studying the quantum dynamics of matter systems interacting with electromagnetic
fields~\cite{Berman1994,Haroche2006}. For a single two-level quantum system
interacting with a single mode of the electromagnetic field of a resonator, a
particularly interesting regime of CQED is reached when the rate, $g$, at which single
excitations are coherently exchanged between the two-level system and the
cavity field mode
exceeds both the decay rate of the two-level system, $\gamma$, and the rate,
$\kappa$, at which the cavity field decays~\cite{Rempe1994Cavity}. This so-called strong coupling regime was
investigated first with single atoms in microwave and optical
cavities~\cite{Brune1996,Thompson1992Observation} and recently with quantum
dots~\cite{Badolato2005Deterministic,khitrova2006} and superconducting Josephson
junctions~\cite{wallraff2004,Chiorescu2004Coherent}. In the optical domain, the use of ultrahigh-finesse
cavities with a very small modevolume allows for reaching the confinement of
the light field required to achieve strong coupling with single neutral
atoms~\cite{Hood1998,Rempe1994Cavity,Maunz2005}. With charged particles, however,
the insertion of dielectric mirrors in the trapping region makes it extremely
challenging to obtain sufficiently small cavity modevolumes, due to the
associated perturbation of the trapping potentials and charging
effects~\cite{Harlander2010Trapped-ion,Herskind2011AMicrofabricated}. 
Although the strong coupling regime has not yet been reached with ions, single
ions in optical cavities have been successfully used for, e.g., probing the
spatial structure of cavity fields~\cite{Guthohrlein2001AsingleIon}, enhanced spectroscopy~\cite{Kreuter2004}, the
generation of single photons~\cite{Keller2004Continuous,Barros2009Deterministic}, the investigation of cavity sideband
cooling~\cite{Leibrandt2009Cavity}, or the demonstration of a single ion laser~\cite{Dubin2010Quantum}. 

For an ensemble of $N$ identical two-level systems simultaneously interacting
with a single mode of the electromagnetic field, the coherent coupling rate is enhanced by a factor
$\sqrt{N}$~\cite{Haroche2006}. This leads to another interesting regime of
CQED, the so-called \textit{collective} strong coupling
regime~\cite{Haroche2006}, where 
the \textit{collective} coupling rate $g_{\mathrm N}=g\sqrt{N}$ is larger than both $\kappa$ and $\gamma$.
This regime, first explored with Rydberg atoms in microwave cavities~\cite{Kaluzny1983}, has been
realized in the optical domain with atomic beams~\cite{Thompson1992Observation}, atoms in magneto-optical
traps~\cite{Lambrecht1996,Nagorny2003Collective,Chan2003Observation,Kruse2004Observation,Chen2011Conditional}, Bose-Einstein
condensates~\cite{Brennecke2007Cavity,Colombe2007Strong}, and, recently, with ion Coulomb
crystals~\cite{Herskind2009Realization}. This cavity-enhanced collective interaction with an
ensemble has many applications within quantum optics and quantum information
processing~\cite{Kimble2008}, including the establishment of strong
nonlinearities~\cite{Lambrecht1995Optical,Joshi2003Optical}, QND
measurements~\cite{Grangier1991Observation,Roch1997Quantum,Mielke1998Nonclassical}, the production~\cite{Black2005On-Demand,Thompson2006AHigh-Brightness}
and storage~\cite{Simon2007,Tanji2009Heralded} of single-photons, the generation of squeezed and entangled
states of light~\cite{Lambrecht1996,Josse2003Polarization,Josse2004Continuous} and atoms
\cite{Leroux2010Implementation,Chen2011Conditional} , the observation of cavity optomechanical
effects~\cite{Nagorny2003Collective,Kruse2004Observation,Klinner2006Normal,Slama2007Superradiant,Murch2008Observation,Brennecke2008}, 
cavity cooling~\cite{Chan2003Observation,Black2003Observation}, and the investigation of quantum phase
transitions~\cite{Baumann2010Dicke}.

This paper provides a detailed description and a theoretical analysis of experiments
achieving \textit{collective} strong coupling with ions~\cite{Herskind2009Realization}. The various
methods used to measure the coherent coupling rate between large ion Coulomb crystals in a linear
quadrupole radiofrequency ion trap and a single field mode of a moderately high-finesse cavity
($\mathcal{F}\sim3000$) are described in detail. Theoretical models based on a semiclassical
approach are applied in assessment of the experimental results of
Ref.~\cite{Herskind2009Realization} as well as of complementary new measurements. Generally, a very good
agreement between the theoretical predictions and the experimental results is
obtained.
As also emphasized in Ref.~\cite{Herskind2009Realization}, the realization of
collective strong coupling with ion crystals is important for ion-based
CQED~\cite{Lange2009CavityQED} and enables, e.g., for the realization of 
quantum information processing devices, such as high-efficiency, long-lived quantum memories~\cite{Lukin2000,Simon2007} and
repeaters~\cite{Duan2001}. In addition to the well-established attractive properties of cold,
trapped ions for quantum information processing~\cite{Leibfried2003Quantum,Blatt2008Entangled}, ion Coulomb
crystals benefit from unique properties which can be exploited for CQED purposes. First, their
uniform density under linear quadrupole trapping conditions~\cite{Drewsen1998,Hornekaer2002Formation,Hornekaer2001} makes it possible to couple the same ensemble equally to different transverse
cavity modes~\cite{Dantan2009Large} and opens for the realization of multimode quantum
light-matter interfaces~\cite{Lvovsky2009Optical}, where the spatial degrees of freedom of light can be exploited in addition
to the traditional polarization and frequency encodings~\cite{Vasilyev2008Quantum,Tordrup2008Holographic,Wesenberg2011Dynamics}. Second, their cold, solid-like nature
combined with their strong optical response to radiation pressure forces and their tunable mode
spectrum~\cite{Dubin1991Theory,Dubin1996Normal,Dantan2010Non-invasive} make ion Coulomb crystals a unique medium to investigate
cavity optomechanical effects~\cite{Kippenberg2008}. Ion Coulomb crystals could, for instance,
be used as a model system to study the back action of the cavity light field on
the collective motion of mesoscopic objects at the quantum limit, as was recently demonstrated with ultracold
atoms~\cite{Murch2008Observation,Brennecke2008,Baumann2010Dicke}. In addition,
novel classical and quantum phase transitions could be investigated using cold
ion Coulomb crystals in optical cavities~\cite{Garcia-Mata2007Frenkel-Kontorova,Retzker2008Double,Fishman2008Structural,Harkonen2009Dicke}.

The paper is organized as follows: Sec.~\ref{sec:theory} presents the theoretical basis for the
CQED interaction of ion Coulomb crystals and an optical cavity field. The cavity field reflectivity
spectra, and the effective number of ions interacting with the cavity field
are derived and the effect of temperature on the collective coupling rate is
discussed. In Sec.~\ref{sec:experimentalSetup} the 
experimental setup and the measurement procedures are described. Section~\ref{sec:experimentalResults}
presents various collective coupling rate measurements and compares them to
the theoretical expectations. Section~\ref{sec:coherencetime} shows measurements of the coherence
time of collective coherences between Zeeman sublevels. A conclusion is given in Sec.~\ref{sec:conclusion}.

\section{CQED interaction: theoretical basis}
\label{sec:theory}
\subsection{Hamiltonian and evolution equations}
\begin{figure}[htb]
  \centering
  \includegraphics[width=\columnwidth]{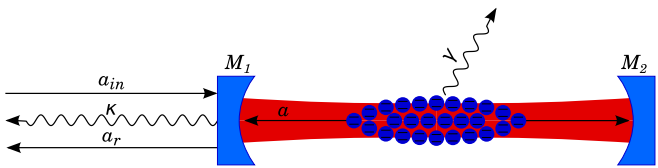}
  \caption[Input-output fields of an optical cavity]{Scheme considered for the description of the
    interaction between an ion Coulomb crystals and the cavity field. The
    cavity is formed by the two mirrors ${M}_{1}$ and
    ${M}_{2}$. $a_{\mathrm{in}}$ is the input light field, $a$ the
    intracavity field, and $a_{\mathrm{r}}$ is the reflected field. $\kappa$
    denotes the cavity field decay rate. The
    spontaneous dipole decay rate of the ions is denoted by $\gamma$. 
  \label{fig:InOutputFieldCavity}}
\end{figure}
We consider the interaction of $N_{\mathrm{tot}}$ two-level ions in a Coulomb crystal with a single mode of
the electromagnetic field of an optical cavity (denoted by $\mathrm{nm}$), as depicted in Fig.~\ref{fig:InOutputFieldCavity}.
The single-ended linear cavity is formed by two mirrors M$_1$ (partial transmitter, PT) and M$_2$
(high reflector, HR) with intensity transmission coefficients $T_1$ and $T_2$
($T_1\gg T_2$). The absorption loss coefficient per round-trip is
$\mathcal{L}$ and the empty cavity field round-trip time is 
$\tau=2l/c$, where $l$ is the cavity length and $c$ the speed of light. The intracavity, input and
reflected fields are denoted by $a$, $a_{\mathrm{in}}$, and $a_{\mathrm{r}}$, respectively. The interaction of an
ensemble of $N$ identical two-level ions with a single mode of the cavity field can be described by
a Jaynes-Cummings Hamiltonian of the form~\cite{Haroche2006,Breuer2007TheTheory}
\begin{equation}
  \label{eq:Jaynes-CummingsHamiltonian}
  H=H_{\mathrm{at}}+H_{\mathrm{l}}+H_{\mathrm{al}}
\end{equation}
where, in the frame rotating at the laser frequency $\omega_{\mathrm{l}}$, the atom and light Hamiltonians are
given by $H_{\mathrm{at}}=\hbar \Delta \sum_{j=1}^{N_{\mathrm{tot}}} \hat \pi_j^{(e)}$ and $H_{\mathrm{l}}=\hbar
\Delta_{\mathrm{c}} \hat a^\dagger \hat a$. The atomic and cavity detunings are denoted by
$\Delta=\omega_{\mathrm{at}}-\omega_{\mathrm{l}}$ and $\Delta_{\mathrm{c}}=\omega_{\mathrm{c}}-\omega_{\mathrm{l}}$, where $\omega_{\mathrm{at}}$ and $\omega_{\mathrm{c}}$
are the atomic and cavity resonance frequencies, respectively. $\hat \pi_j^{(e)}$ is the excited
state population operator of the $j$-th ion and $\hat{a}$, $\hat{a}^{\dagger}$ are the intracavity field
annihilation and creation operators. In the rotating wave approximation the interaction Hamiltonian
reads
\begin{equation}
  \label{eq:interactionHamiltonian}
  H_{\mathrm{al}}=-\hbar g\sum_{j=1}^{N_{\mathrm{tot}}} \Psi_{\mathrm{nm}}(\bm r_j) (\hat\sigma_j^\dagger \hat a +\hat\sigma_j \hat a^\dagger).
\end{equation}
where $\hat \sigma_j^\dagger$ and $\hat \sigma_j$ are the atomic rising and lowering operators,
defined in the frame rotating at the laser frequency. The single-ion coupling rate $g$ is defined
as $g=\mu_{\mathrm{ge}}E_{0}/\hbar$, where $\mu_{\mathrm{ge}}$ is the dipole element of the transition considered and
$E_0$ the maximum electric field amplitude. The field distribution $E_0
\Psi_{\mathrm{nm}}(\bm r_j)$ is assumed to be
that of a single-cavity Hermite-Gauss mode~\cite{Kogelnik1966Laser}. In the following, we will restrict
ourselves to the fundamental $\mathrm{TEM}_{00}$ mode of the cavity and refer to
Ref.~\cite{Dantan2009Large} for the coupling of ion Coulomb crystals to higher-order cavity
transverse modes.

The coupled atom-cavity system is subject to decoherence, mainly through the spontaneous decay of the
ions from the excited state and through the decay of the cavity field due to the finite
reflectivity of the cavity mirrors and due to intracavity losses. 
These dissipative processes are characterized by the atomic
dipole decay rate, $\gamma$, and by the total cavity field decay
rate, $\kappa$, respectively. The cavity field decay rate is given by
$\kappa=\kappa_1+\kappa_2+\kappa_{\mathcal{L}}$, and includes the decay rates
through the PT and HR mirrors ($\kappa_1=T_1/2\tau$ and $\kappa_2=T_2/2\tau$) and the decay rate due to absorption losses
($\kappa_{\mathcal{L}}=\mathcal{L}/2\tau$).

We derive standard semiclassical equations of motion for the mean values of
the observables via $\langle \dot{\hat{a}}\rangle=\frac {i}{\hbar}\langle[H,\hat{a}]\rangle$
and phenomenologically adding the relevant dissipative processes \cite{Haroche2006,Breuer2007TheTheory,Scully1997,Tavis1968Exact,Thompson1992Observation,Raizen1989Normal-mode}.

In the low saturation
regime, most of the atoms remain in the ground state, $\langle
\hat{\pi}_j^{(e)}\rangle\ll 1$, and the dynamical equations for the mean
values of the observables read 
\begin{eqnarray}
  \label{eq:opticalBlochEquationsLowSaturation1}
  \dot \sigma_j &=& -(\gamma+i \Delta) \sigma_j + i g \Psi_{00}(\bm r_j) a, \\
  \dot a &=& -(\kappa+i \Delta_{\mathrm{c}}) a + i \sum_{j=1}^{N_{\mathrm{tot}}} g
  \Psi_{00}(\bm r_j) \sigma_j +\sqrt{2\kappa_1/\tau} a_{\mathrm{in}}.\label{eq:opticalBlochEquationsLowSaturation2}
\end{eqnarray}
where $o=\langle \hat o\rangle$ is the mean value of observable $\hat o$.

\subsection{Steady-state reflectivity spectrum and effective number of ions}
\begin{figure}[tb]
  \centering
  \includegraphics[width=\columnwidth]{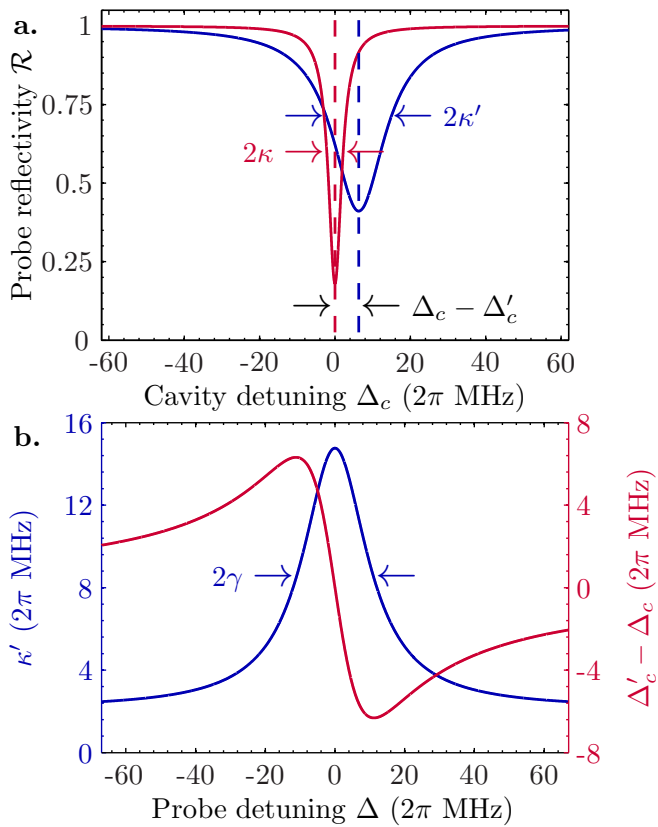}
  \caption{(Color online) (a) Calculated probe reflectivity spectrum as a
    function of cavity detuning $\Delta_{\mathrm c}$ for an empty cavity (red
    line) and for $N=500$ ions interacting with the cavity field (blue line)
    [see Eq.~\eqref{eq:reflectivity}] for a probe detuning of
    $\Delta=\gamma$. The coupling of the ions and the cavity mode lead to a
    broadening and a shift of the resonance dip. The values of the parameters 
    are set to typical values used in the experiments presented in
    Secs. \ref{sec:experimentalSetup} and \ref{sec:experimentalResults}: $\kappa=2\pi\times 2.1~\mathrm{MHz}$, 
    $\kappa_1=2\pi\times1.5~\mathrm{MHz}$, $\gamma=2\pi\times
    11.2~\mathrm{MHz}$, $g=2\pi\times 0.53~\mathrm{MHz}$.  
    (b) Expected effective cavity decay rate $\kappa^\prime$ and shift of the 
    cavity resonance $\Delta^\prime_{\mathrm{c}}-\Delta_{\mathrm{c}}$ as a
    function of probe detuning $\Delta$ calculated for the same set of
    parameters. 
    \label{fig:theoryCoupledAtomCavitySystem}}
\end{figure}

In steady-state, the mean value of the intracavity field amplitude is given by
\begin{eqnarray}
  \label{eq:steadyStateSolution}
  a=\frac {\sqrt{2 \kappa_1/\tau}a_{\mathrm{in}}}{\kappa^\prime+i\Delta_{\mathrm{c}}^\prime},
\end{eqnarray}
where an effective cavity decay rate and an effective cavity detuning are introduced:
\begin{eqnarray}
  \label{eq:kappaPrime}
  \kappa^\prime&=&\kappa+g^2 N\frac{\gamma}{\gamma^2+\Delta^2},\\
  \Delta_{\mathrm{c}}^\prime&=&\Delta_{\mathrm{c}} - g^2 N\frac{\Delta}{\gamma^2+\Delta^2}.\label{eq:delta_cprime}
\end{eqnarray}
In these expressions, $N$ is the effective number of ions interacting with the
intracavity field, which is calculated by summing over all ions and weighting
the contribution of each ion by the field modefunction under consideration
evaluated at the ion's position:
\begin{equation}
  \label{eq:effectiveNumberOfIons}
  N= \sum_{j=1}^{N_\mathrm{tot}} \Psi_{\mathrm{00}}^2(\bm r_j).
\end{equation}
Here,
\begin{align}\nonumber\Psi_{00}^2(\bm r)=&\left(\frac{w_0}{w(z)}\right)^2\exp\left(-\frac{2r^2}{w(z)^2}\right)\\
&\times\sin^2\left[kz-\arctan(z/z_0)+kr^2/2R(z)\right],\end{align} is the modefunction of the cavity
fundamental TEM$_{00}$ Gaussian mode with waist $w_0$ at the center of the mode and
\hbox{$w(z)=w_0\sqrt{1+z^2/z_0^2}$}, \hbox{$R(z)=z+z_0^2/z$}, \hbox{$z_0=\pi
  w_0^2/\lambda$}, and \hbox{$k=2\pi /\lambda$}.

Large ion Coulomb crystals in a linear radiofrequency trap are to an excellent approximation
spheroids with half-length $L$ and radius $R$ (see
Fig.~\ref{fig:crystalAbsorptionAndPhaseShift}), where the density of ions,
$\rho$, is constant
throughout the crystal~\cite{Drewsen1998,Hornekaer2002Formation}. It is then convenient to adopt a
continuous medium description, in which Eq.~(\ref{eq:effectiveNumberOfIons}) becomes an integral
over the crystal volume $V$:
\begin{equation}
  N=\rho\int_V d\bm r\Psi_{00}^2(\bm r)
\end{equation} 
In our experiment, the crystal radius and half-length, $R$ and $L$, are
typically much smaller than $z_0$ and the axial mode function can be
approximated by $\sin^2(kz)$. Moreover, for randomly distributed ions along
the cavity axis $z$, one can average over the cavity standing-wave
longitudinal structure, which gives an effective number of ions equal to 
\begin{equation}
  \label{eq:effectiveNumberOfIonsAxiallyAveraged}
  N=\frac{\rho}{2}\int_V d\bm r \exp[-2r^2/w(z)^2].
\end{equation} 
This expression can be evaluated knowing the crystal
dimensions, its density, and the cavity mode geometry. For typical crystals
with large radial extension as compared to the cavity waist $R\gg w_0$ and
length smaller than the Rayleigh range $L\ll z_{0}$, this expression reduces
to \begin{equation}N\simeq\rho\frac{\pi w_0^2}{4}L,\end{equation} which is
simply the product of the ion density by the volume of the cavity mode in the
crystal. 

Using the input-output relation $a_{\mathrm{r}}=\sqrt{2\kappa_1
  \tau}a-a_{\mathrm{in}}$, one finds that the steady-state probe reflectivity
spectrum of the cavity is also Lorentzian-shaped in 
presence of the ions, the bare cavity decay rate, and detuning $\kappa$ and
$\Delta_{\mathrm{c}}$ being replaced by their effective 
counterparts $\kappa'$ and $\Delta_{\mathrm{c}}'$ of Eqs.~(\ref{eq:kappaPrime},\ref{eq:delta_cprime}):
\begin{equation}
  \label{eq:reflectivity}
  \mathcal{R}\equiv\left|\frac{a_r}{a_{\mathrm{in}}}\right|^2=
  \left|\frac{2\kappa_1-\kappa'-i\Delta_{\mathrm{c}}'}{\kappa'+i\Delta_{\mathrm{c}}'}\right|^2.
\end{equation}
The broadening and shift of the cavity resonance then represent the change in absorption and
dispersion experienced by the cavity field interacting with $N$ ions. In
Fig. \ref{fig:theoryCoupledAtomCavitySystem} (a) the expected cavity
reflectivity spectrum is shown for both an empty
cavity and a cavity containing a crystal with an effective number of ions
$N=500$ and for parameters corresponding to those used in the experiments
presented in Secs. \ref{sec:experimentalSetup} and \ref{sec:experimentalResults}.
In Fig. \ref{fig:theoryCoupledAtomCavitySystem} (b) the effective
cavity decay rate, $\kappa^\prime$, and the shift of the 
cavity resonance induced by the interaction with the ions,
$\Delta^\prime_{\mathrm{c}}-\Delta_{\mathrm{c}}$, are shown as a function of
the probe detuning, $\Delta$, for the same parameters.

\subsection{Effect of the motion of the ions}
\label{sec:EffectOfTheMotionOfTheIons}
The interaction Hamiltonian in Eq.~\eqref{eq:interactionHamiltonian} is only valid for atoms at
rest. If an ion is moving along the axis of the cavity, the standing-wave structure of the cavity
field and the Doppler shifts due to the finite velocity of the ion have to be taken into account.
For an ion moving along the standing wave field with a velocity $v_j$, it is convenient to define
atomic dipole operators, $\sigma_{j\pm}=\frac{1}{2}\sigma_j \exp{(\pm i k z_j)}$, arising from the
interaction with the two counterpropagating components of the standing-wave cavity field. In the
low saturation limit and taking into account the opposite Doppler-shifts, the evolution equations
\eqref{eq:opticalBlochEquationsLowSaturation1},\eqref{eq:opticalBlochEquationsLowSaturation2}
become
\begin{widetext}
\begin{eqnarray}
  \label{eq:opticalBlochEquationsIncludingMotion}
  \dot \sigma_{j\pm} &=& -\left[\gamma+i (\Delta\pm
    kv_j)\right]\sigma_{j\pm}+i (g/2) \Psi_{\mathrm{nm}}(\bm r_j) a\\
  \dot a &=& -(\kappa+i \Delta_{\mathrm{c}}) a + i (g/2) \sum_{j=1}^{N_{tot}} \Psi_{\mathrm{nm}}(\bm r_j) \left(\sigma_{j+}+\sigma_{j-}\right)+\sqrt{2\kappa_1/\tau} a_{\mathrm{in}}.
\end{eqnarray}
\end{widetext}
When the typical timescale of the motion is slow as compared to the timescales for the coupled
dynamics of the atomic dipole and cavity field, the steady-state mean value of the intracavity
field can be found by averaging the contributions of the individual dipole mean values given by
Eq.~\eqref{eq:opticalBlochEquationsIncludingMotion} over the distribution of
the mean velocities, $f(v)$. For a distribution $f(v)$ with an average
velocity $v_{\mathrm{D}}$ a conservative estimate for this to be valid is that the mean Doppler-shift is
smaller than both effective rates of the coupled system on resonance
($\Delta_{\mathrm{c}}=\Delta=0$), $kv_{\mathrm{D}}\ll\min[\kappa+g^2N/\gamma,\gamma+g^2N/\kappa]$. Under these conditions,
the expression for the intracavity field mean value is then of the same form as in the
zero-velocity case (Eq.~\eqref{eq:steadyStateSolution}). The effective cavity field decay rate and detuning of
Eqs.~\eqref{eq:kappaPrime} and\eqref{eq:delta_cprime} are modified according to
\begin{eqnarray}
  \label{eq:kappaMotion}
 \kappa^\prime&=&\kappa+g^2 N \int {\mathrm d} v f(v)\gamma \xi(v)  \\
 \label{eq:DeltaCMotion}\Delta_{\mathrm{c}}^\prime&=&\Delta_{\mathrm{c}}-g^2 N \int {\mathrm d} v f(v)(\Delta-kv) \xi(v) .
\end{eqnarray}
where \begin{equation}\xi(v)=\frac{\gamma ^2 + \Delta ^2 +(kv)^2}{(\gamma^2 + \Delta ^2)^2
+2(\gamma^2 - \Delta ^2)(kv)^2 + (kv)^4}.\end{equation} In the case of a thermal Maxwell-Boltzmann
distribution with temperature $T$, one has \hbox{$ f(v) = \sqrt{\frac m {2 \pi
k_{\mathrm{B}}T}}\exp\left(-\frac {m v^2}{2k_{\mathrm{B}}T}\right)$}, where $k_{\mathrm{B}}$ is the Boltzmann
constant and $m$ the mass of the ion. At low temperatures, i.e., when the width of the thermal
distribution is small as compared to the atomic natural linewidth, the effective cavity field decay
rate and detuning given by Eqs.~\eqref{eq:kappaMotion} and \eqref{eq:DeltaCMotion} are
well-approximated by 
\begin{eqnarray}
  \label{eq:kappaPrimeTemperature}
  \kappa^\prime&=&\kappa+g^2 N\frac{\gamma^\prime}{{\gamma^\prime}^2+\Delta^2},\\
  \Delta_{\mathrm{c}}^\prime&=&\Delta_{\mathrm{c}} - g^2 N\frac{\Delta}{{\gamma^\prime}^2+\Delta^2}.\label{eq:delta_cprime_Temperature}
\end{eqnarray}
These equations are of the same form as
Eqs.~\eqref{eq:kappaPrime} and \eqref{eq:delta_cprime}, replacing the natural
dipole decay rate by an effective dipole decay rate  ,
\begin{equation}
  \label{eq:effectiveGamma}
  \gamma^\prime\simeq \gamma(1+kv_{\mathrm{D}}/\sqrt{2}),
\end{equation}
where $v_{\mathrm{D}}=\sqrt{k_BT/m}$ is the mean Doppler velocity.

\section{Experimental setup}
\label{sec:experimentalSetup}
\begin{figure}[h!]
  \includegraphics[width=\columnwidth]{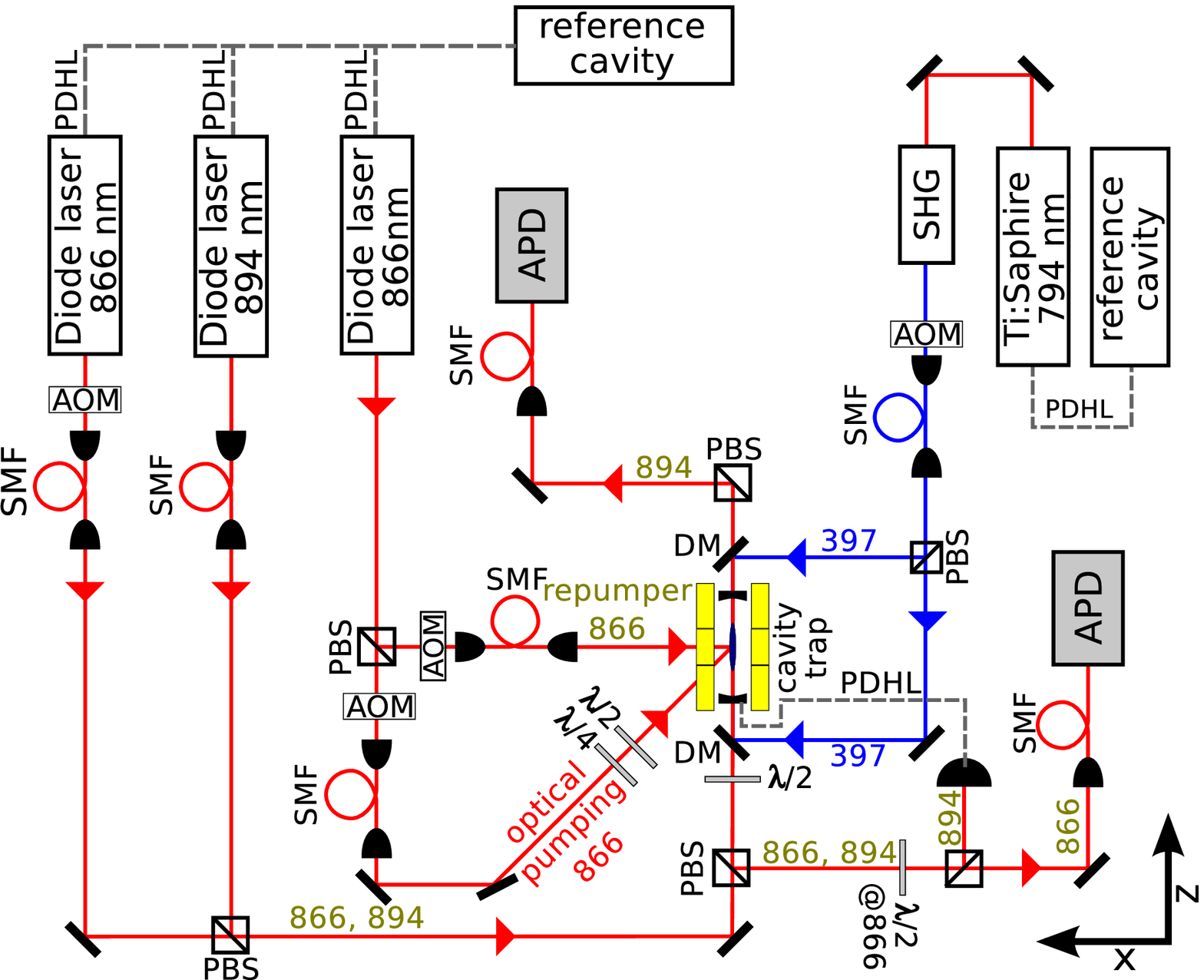}
  \caption{(Color online) Schematic experimental setup. The abbreviations
    are: polarizing beam splitter (PBS), single mode fiber (SMF),
      acousto-optical modulator (AOM), dichroic mirror (DM), Pound-Drever-Hall
      lock (PDHL), avalanche photodiode (APD), second harmonic generation (SHG). The photoionization laser is not shown.}
\label{fig:ExperimentalSetup}
\end{figure}
\subsection{Cavity trap}
The ion trap used is a segmented linear quadrupole radiofrequency trap that
consists of four cylindrical electrode rods (for details see \cite{Herskind2008Loading}). The electrode radius is
$2.60~\mathrm{mm}$ and the distance from the trap center to the electrodes is $r_0 = 2.35~\mathrm{mm}$. Each electrode rod is
divided into three parts, where the length of the center electrode is
$z_{\mathrm{C}}=5.0~\mathrm{mm}$, and the length of the end electrodes is
$z_{\mathrm{E}}=5.9~\mathrm{mm}$. Radial confinement is achieved by a radiofrequency
field (RF) applied to the entire rods at a frequency of $2\pi \times
4~\mathrm{MHz}$ and a $\pi$ phase difference between neighboring rods. The
axial trapping potential is created by static voltages (DC) applied to the 
outer parts of the rods.

An optical cavity is incorporated into the trap with its axis parallel to the symmetry axis of the
ion trap (see Fig. \ref{fig:ExperimentalSetup}). The cavity mirrors have a diameter of $1.2~\mathrm{mm}$ and a radius of curvature of
$10~\mathrm{mm}$. The rear face of both mirrors are anti-reflection coated at a wavelength of $866~\mathrm{nm}$
corresponding to the $3\mathrm{d}^2\mathrm{D}_{3/2}\leftrightarrow 4\mathrm{p}^2\mathrm{P}_{1/2}$
transition in $^{40}\mathrm{Ca}^+$, while the front facade of one mirror is
partially transmitting (PT) and for the other
highly reflecting (HR) at this wavelength. Their intensity transmission coefficients are $1500$ and
$5~\mathrm{ppm}$, respectively. The intracavity losses due to
contamination of the mirrors during the initial bake out amount to $\sim 650$ ppm. The PT mirror is mounted on a plate that
can be translated using piezoelectric actuators to allow for scanning or actively stabilizing the
cavity length. The cavity has a close to confocal geometry with a length of $11.8~\mathrm{mm}$,
corresponding to a free spectral range of $12.7~\mathrm{GHz}$ and a waist of the fundamental
$\mathrm{TEM_{00}}$ mode of $w_0=37 \mathrm{\mu m}$. With a measured cavity field decay rate of
$\kappa=2\pi \times (2.1\pm0.1)~\mathrm{MHz}$, the finesse is found to be $\mathcal{F}=3000\pm200$
at a wavelength of $866~\mathrm{nm}$~\cite{Herskind2008Loading}.

$^{40}\mathrm{Ca}^+$ ions are loaded into the trap by \textit{in situ} photoionization of atoms from a beam of
atomic calcium in a two-photon resonant photoionization process~\cite{Kjaergaard2000Isotope,Mortensen2004Isotope,Herskind2008Loading}. The
ions are cooled to a crystalline state through Doppler-laser cooling using a combination of two
counterpropagating laser beams, resonant with the $4\mathrm{s}^2\mathrm{S}_{1/2}\leftrightarrow
4\mathrm{p}^2\mathrm{P}_{1/2}$ transition at 397 nm along the trap axis, and a repumping laser
applied along the $x$ axis and resonant with the $3\mathrm{d}^2\mathrm{D}_{3/2}\leftrightarrow
4\mathrm{p}^2\mathrm{P}_{1/2}$ transition at 866 nm to prevent shelving to the metastable
$\mathrm{D}_{3/2}$ state. Three sets of Helmholtz coils are used to compensate for residual
magnetic fields and to produce bias magnetic fields. For the measurements of the collective
coupling rate between the ion Coulomb crystals and the cavity light field, the transverse magnetic fields
along $x$ and $y$ are nulled and a magnetic field of $B_{\mathrm{z}}\sim 2.5~{G}$ along the $z$-axis is
used.

\subsection{Detection}
A grating stabilized diode laser at $866~\mathrm{nm}$ provides the light for probing the coupling of
the ion Coulomb crystals with the standing wave field inside the optical
cavity. It is injected into the cavity through the PT mirror.
Additionally, a second grating stabilized diode laser with a wavelength of $894~\mathrm{nm}$
serves as an off-resonant reference laser and is simultaneously coupled to
the cavity through the PT mirror and used to monitor the cavity resonance. Both lasers are
frequency stabilized to the same temperature stabilized reference cavity and have linewidths of $\sim
100~\mathrm{kHz}$.

The reflectivity of the 866-nm cavity field is measured using an avalanche photo diode (APD). The
light sent to the APD is spectrally filtered by a diffraction grating ($1800~\mathrm{lines/mm}$)
and coupled to a single mode fiber. Taking into account the efficiency of the
APD at $866~\mathrm{nm}$, the fiber incoupling and the optical losses, the
total detection efficiency amounts to $\approx 16\%$. A similar detection system is used to measure the
transmission of the 894-nm reference laser.

Depending on the experiment, the reference laser serves two different purposes. In a first
configuration, the length of the cavity is scanned at a rate of $30~\mathrm{Hz}$ over the atomic
resonance. In this configuration, the frequency of the reference laser is tuned such that it is
resonant at the same time as the probe laser in the cavity scan. This allows for monitoring slow
drifts and acoustic vibrations. The signal of the weak probe laser is then averaged over typically 100
scans in which the stronger reference laser is used to keep track of the
current position of the cavity resonance.

In a second configuration, the cavity resonance is locked on the atomic resonance by stabilizing the
length of the cavity to the frequency of the reference laser in a Pound-Drever-Hall locking scheme~\cite{Drever1983Laser}.
During the measurement, imperfections in the stabilization are compensated for by monitoring the
transmission of the 894-nm reference laser. The data is then postselected by only
keeping data points for which the transmitted reference signal was above a certain threshold.

\subsection{Experimental sequence}
\label{sec:experimentalSequence}
\begin{figure*}
  \includegraphics[width=1.5\columnwidth]{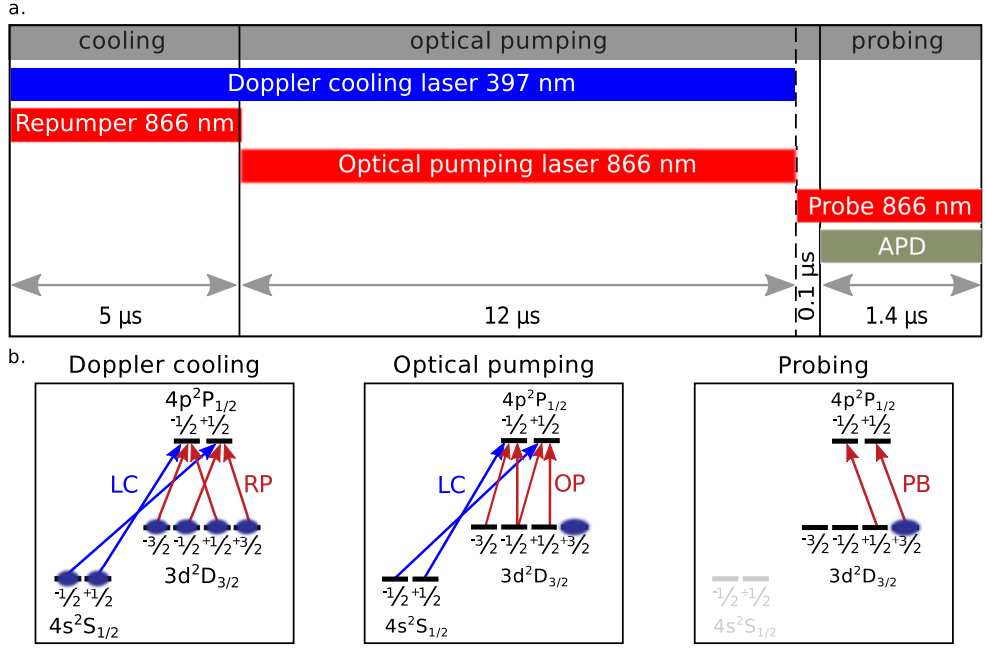}
    \caption{(color online) (a) Experimental sequence used to measure the collective coupling rate. (b) Energy levels of $^{40}\mathrm{Ca}^+$ including
      the relevant transitions addressed in the three parts of
      the experimental sequence. The acronyms are: laser cooling beam (LC),
      repumping beam (RP), optical pumping beam (OP), probe beam (PB),
      avalanche photodiode (APD). 
    \label{fig:experimentalSequence}}
\end{figure*}
In both configurations, the cavity reflection spectrum is measured at a rate of $50~\mathrm{kHz}$
using a $20~\mathrm{\mu s}$ sequence of Doppler cooling, optical pumping and probing, as indicated
in Fig.~\ref{fig:experimentalSequence}. 
First, the ions are Doppler-laser cooled for $5~\mathrm{\mu s}$ by driving the $4\mathrm{s}^2\mathrm{S}_{1/2}\leftrightarrow 4\mathrm{p}^2\mathrm{P}_{1/2}$
transition using laser cooling beams at 397~nm (LC), and at the same time repumping on the $3\mathrm{d}^2\mathrm{D}_{3/2}\leftrightarrow
4\mathrm{p}^2\mathrm{P}_{1/2}$ transition with a laser at 866~nm (RP).
Next, the ions are optically pumped to the ${m_{\mathrm{J}}}=+3/2$ magnetic substate of the
$3\mathrm{d}^2\mathrm{D}_{3/2}$ level by applying the optical
pumping laser (OP) in combination with the laser
cooling beams (LC) for a period of $12~\mathrm{\mu s}$. The optical pumping
laser is resonant with the $3\mathrm{d}^2\mathrm{D}_{3/2}\leftrightarrow
4\mathrm{p}^2\mathrm{P}_{1/2}$ transition and has a polarization consisting only of
$\sigma^+$- and $\pi$-polarized components. It is sent to the trap under an
angle of $45^\circ$ with respect to the quantization axis. By probing the
populations of the different Zeeman sublevels, the efficiency of the optical
pumping was measured to be $\eta = 97^{+3}_{-5}\%$~\cite{phdPeterHerskind}. 
Finally, the cavity
reflection signal is probed by injecting a $1.4~\mathrm{\mu s}$ $\sigma^-$-polarized probe pulse,
resonant with the $3\mathrm{d}^2\mathrm{D}_{3/2}\leftrightarrow 4\mathrm{p}^2\mathrm{P}_{1/2}$
transition, into the $\mathrm{TEM}_{00}$ mode of the optical cavity. Its intensity is set such that
the mean intracavity photon number is less than one at any time. With a delay of $0.1~\mathrm{\mu
s}$ relative to the probe laser, the APD is turned on. The delay ensures that the field has built
up inside the cavity and that the system has reached a quasi-steady-state. The length of the
probing period was chosen in order to minimize the total sequence length as well as to avoid
depopulation due to saturation of the transition~\cite{phdPeterHerskind}.

\subsection{Effective number of ions}\label{sec:effectivenumberofions}
\begin{figure}[htb]
  \includegraphics[width=\columnwidth]{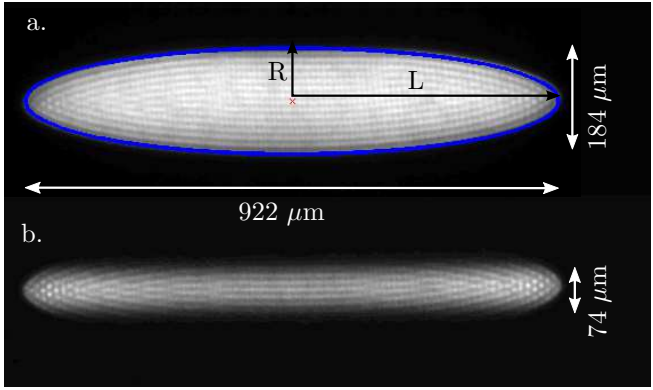}
  \caption[Crystal image for absorption measurement]{(Color online) (a) Typical projection image of a crystal used in collective strong coupling
  measurements. All ions are exposed to cooling and repumping light. The solid
  blue line indicates the outline of the ellipsoid. 
  (b) Same crystal, but only the ions in the cavity mode are exposed to repumping light, now injected into the cavity. The
      ions outside the cavity mode volume are not visible, as they are shelved into the metastable
      $3\mathrm{d}^2\mathrm{D}_{3/2}$ level. The crystal contains $N_{\mathrm{tot}}=8780\pm180$ ions, of which
      %$N=504\pm10$ 
      $N=489^{+18}_{-27}$ effectively interact with the cavity field.
    \label{fig:crystalAbsorptionAndPhaseShift}}
\end{figure}
As mentioned above, the effective number of ions interacting with the cavity field depends on the ion
crystal density and the overlap between the crystal and the cavity modevolume,
where the density of the ion Coulomb crystals depends on the amplitude of the
RF voltage~\cite{Hornekaer2001}:
\begin{equation}
  \label{eq:ionDensity}
  \rho=\frac {\epsilon_0 U_{\mathrm{RF}}^2}{M r_0^4 \Omega_{\mathrm{RF}}^2}.
\end{equation}
Here, $M$ denotes the ion mass. The precise calibration of the RF voltage on
the trap electrodes can be performed, e.g., on the basis of a zero-temperature
charged liquid
model~\cite{Turner1987Collective,Hornekaer2001,Herskind2009Positioning} or 
the measurement of the Wigner-Seitz radius~\cite{Herskind2009Positioning}. For the trap used in
these experiments, $\rho=(6.01\pm0.08)\times 10^3 ~U_{\mathrm{RF}}^2~\mathrm{V^{-2}cm^{-3}}$. The
crystal mode volume is found by taking fluorescence images of the crystal
during Doppler-laser cooling, as shown
in Fig.~\ref{fig:crystalAbsorptionAndPhaseShift}, from which the crystal half-length $L$ and radius
$R$ can be extracted. Taking a possible offset between the cavity axis and the
crystal revolution axis into account, the effective number of ions [see
Eq.~\eqref{eq:effectiveNumberOfIonsAxiallyAveraged}] is then numerically
calculated using the formula  
\begin{equation}
  N=\eta \frac { \rho} 2 \int_V \mathrm{d}x
  \mathrm{d}y\;\exp \left\{ -2[(x-x_0)^2+(y-y_0)^2]/w_0^2\right\},
\end{equation}
where the parameter $\eta$ accounts for a finite efficiency of the optical
pumping preparation and $x_0$ and $y_0$ denote the radial offsets. These offsets can in principle be canceled
to within a micron~\cite{Herskind2009Positioning}, but in the experiments reported here, they were
measured to be $x_0=3.9~\mathrm{\mu m},~y_0=15.7~\mathrm{\mu m}$~\cite{Dantan2009Large}. The
uncertainty in the effective number of ions comes from the uncertainty $\delta \rho$ in the
density determination, due to the RF voltage calibration, the uncertainty in the crystal volume
$\delta V$, due to the imaging resolution $\delta x$ and the uncertainty of
the optical pumping efficiency $\delta \eta$. The relative uncertainty in the
effective number of
ions, $N=\eta \rho V$, can then be expressed as~\cite{phdPeterHerskind}
\begin{equation}
  \label{eq:errorN}
  \frac{\delta N}{N}=\sqrt{\left(\frac{\delta\rho}{\rho}\right)^2+\left(\frac{\delta
        V}{V}\right)^2+\left(\frac {\delta \eta}{\eta}\right)^2},
\end{equation} 
where $\delta V/V=\delta x\sqrt{16L^2+R^2}/2RL$. For the typically
few-mm-long prolate crystals used in these experiments and an imaging resolution $\delta x\sim $
$\mu$m, this results in a relative uncertainty of 5-7\% in the effective number of ions.

\section{Collective coupling measurements}
\label{sec:experimentalResults}

To achieve collective strong coupling on the chosen
$3\mathrm{d}^2\mathrm{D}_{3/2},\;m_{\mathrm{J}}=+3/2\leftrightarrow 4\mathrm{p}^2\mathrm{P}_{1/2},\;m_{\mathrm{J}}=+1/2$
transition the collective coupling rate $g\sqrt{N}$ has to be larger than the cavity
field decay rate $\kappa=2\pi \times 2.1$ MHz and the optical dipole decay rate $\gamma=2\pi\times11.2$ MHz.
With the known dipole element of the transition and the cavity geometry, the single-ion
coupling rate at an antinode at the center of the cavity fundamental mode is
expected to be $g=2\pi\times(0.53\pm0.01)$~MHz. One thus expects to
be able to operate in the collective strong coupling regime as soon as
$N\gtrsim 500$. 

\subsection{Atomic absorption and dispersion}
\label{sec:CollectiveStrongCoupling}

To investigate the coherent coupling of the ions with the cavity field in the collective strong
coupling regime, we first perform measurements of the atomic absorption and dispersion of a given
crystal with $N\sim 500$ by scanning the cavity length around atomic resonance and
recording the probe reflectivity spectrum. The crystal used in these experiments is similar to the one
shown in Fig.~\ref{fig:crystalAbsorptionAndPhaseShift}. With a density of
$\rho=(5.4\pm0.1)\times10^8~\mathrm{cm}^{-3}$ , a half-length $L=(511\pm1)~\mathrm{\mu m}$ and radius
$R=(75\pm1)~\mathrm{\mu m}$ the total number of ions in the crystal is
$N_{\mathrm{tot}}=6500\pm200$, and the effective number of ions interacting
with the cavity mode is $N=520^{+24}_{-32}$.
  \begin{figure}
    \centering
    \includegraphics[width=\columnwidth]{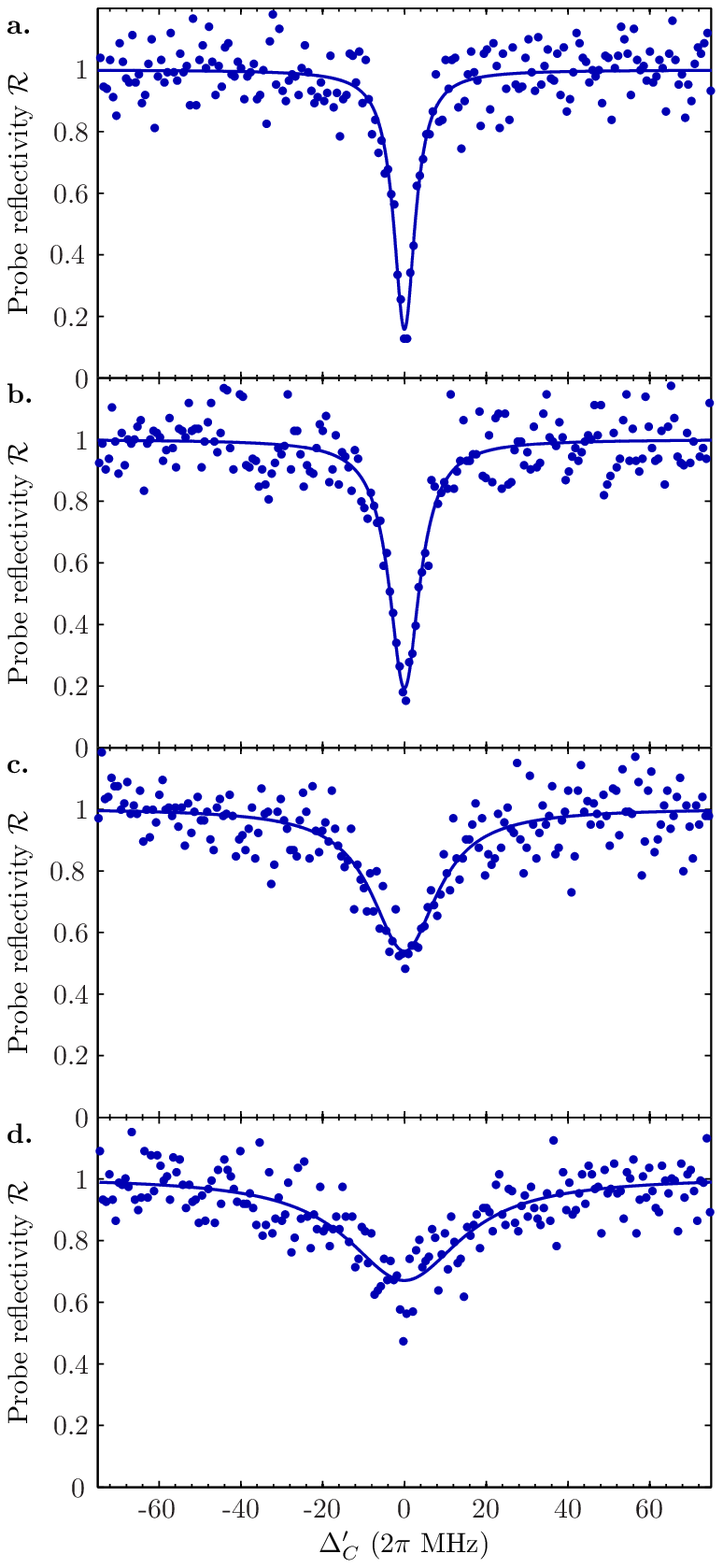}
    \caption{(Color online) Typical probe reflectivity for various values of the
      atomic detuning $\Delta$. The probe detunings were (a) $\Delta\approx
      2\pi\times54.3~\mathrm{MHz}$, (b) $\Delta\approx 2\pi\times24.3~\mathrm{MHz}$,
      (c.) $\Delta\approx 2\pi\times8.3~\mathrm{MHz}$, and (d.) $\Delta\approx
      2\pi\times0.3~\mathrm{MHz}$. Solid lines are Lorentzian fits to the data; the
      effective cavity field decay rate $\kappa'$ is deduced from the fit.
  \label{fig:Abs_scans_all}}
\end{figure}
\begin{figure}
  \includegraphics[width=\columnwidth]{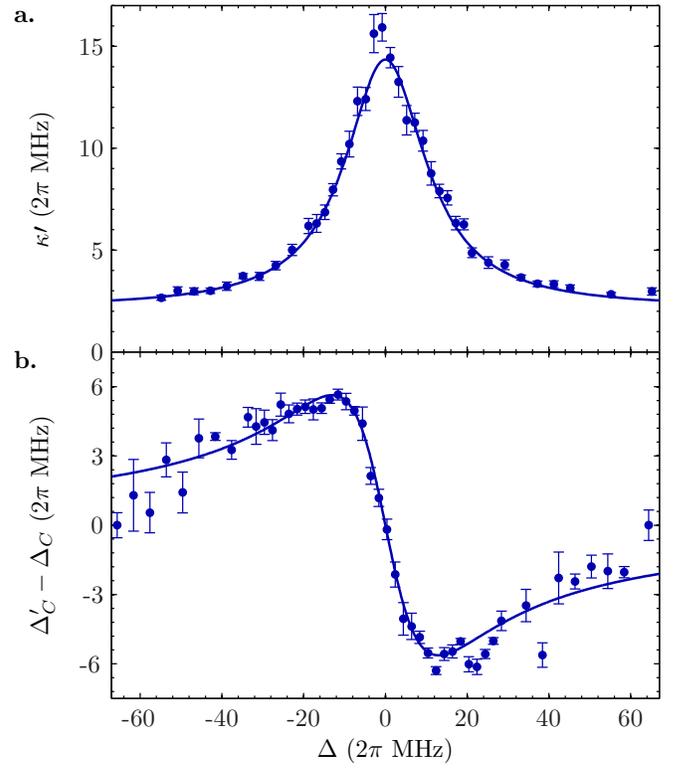}
  \caption{\label{fig:absorptionWidthAndPhaseShift}(Color online)
    (a) Measured cavity field effective decay rate $\kappa'$ versus probe detuning $\Delta$
    for a crystal with $N=520^{+24}_{-32}$ ions interacting with the cavity
    field. The blue solid line is a fit to the data.
    (b) Measured shift of the cavity resonance frequency $\Delta_{\mathrm{c}}'-\Delta_{\mathrm{c}}$ versus atomic
    detuning $\Delta$ for the same crystal. The blue line is a fit to the data.}
\end{figure}

The broadening and the shift of the cavity resonance are then measured as a function of the
detuning of the probe laser, $\Delta$. This is accomplished by scanning the cavity length over a
range corresponding to $\sim1.3~\mathrm{GHz}$ at a repetition rate of $30~\mathrm{Hz}$, for a fixed
value of $\Delta$. The width of the reflection dip for a given detuning $\Delta$ is found by
averaging over 100 cavity scans, where the reference laser is overlapped with the probe laser on
the cavity scan and used to compensate for any drift of the cavity. In Fig.~\ref{fig:Abs_scans_all},
cavity reflection scans are plotted for various detunings. Each data point corresponds to
the average of 100 $20-\mathrm{\mu s}$-measurement sequences as showed in
Fig.~\ref{fig:experimentalSequence}. As expected from Eq.~\eqref{eq:kappaPrimeTemperature}, the broadening of
the intracavity field absorption reflects the two-level atomic medium absorption. Each set of data
is, according to Eq.~\eqref{eq:reflectivity}, fitted to a Lorentzian from which
the cavity half width half maximum (HWHM) $\kappa'$ is deduced. Figure 
\ref{fig:absorptionWidthAndPhaseShift}(a) shows the modified cavity HWHM, $\kappa^\prime$, as a
function of detuning of the probe laser, $\Delta$. Each point is the average of five measurements; the
solid line is a fit according to Eq.~\eqref{eq:kappaPrimeTemperature}. From the fit we deduce a collective
coupling rate of $g_{\mathrm{N}}=2\pi\times(12.2\pm0.2)~\mathrm{MHz}$, in good agreement with the theoretical
expectation of $g_{\mathrm{N,~theory}}=2\pi\times(12.1^{+0.4}_{-0.5})~\mathrm{MHz}$,
calculated for $N=520^{+24}_{-32}$ ions interacting with the cavity mode~\cite{Herskind2009Realization}. Furthermore, the effective dipole decay rate
$\gamma'$ is left as a fit parameter to account for nonzero temperature
effects, as discussed in
Sec.~\ref{sec:EffectOfTheMotionOfTheIons}. %Eq.~\eqref{eq:kappaMotion}. 
The fit yields $\gamma^\prime=2\pi\times(11.9\pm0.4)~\mathrm{MHz}$, which would
correspond to a temperature of $T=24^{+20}_{-14}~\mathrm{mK}$, and a natural half-width of
the cavity of $\kappa=2\pi\times(2.2\pm0.1)~\mathrm{MHz}$, in good agreement with the value deduced
from an independent measurement of the free spectral range (FSR) and the finesse of the cavity,
$\kappa=2\pi\times(2.1\pm0.1)~\mathrm{MHz}$~\cite{Herskind2008Loading}.

For the measurement of the effective cavity detuning, $\Delta_{\mathrm{c}}'$, the position of the
894-nm resonance laser in the cavity scan is fixed to the bare cavity resonance. The
frequency shift is then measured by comparing the position of the probe and the reference signal
resonances in the cavity scan. The effective cavity detuning as a function of probe detuning is
shown in Fig.~\ref{fig:absorptionWidthAndPhaseShift}(b) One observes the typical dispersive
frequency-shift of two-level atoms probed in the low saturation regime. The data is fitted
to the theoretical model according to Eq.~\eqref{eq:delta_cprime_Temperature}, to find a
collective coupling rate 
$g_{\mathrm{N}}=2\pi\times(12.0\pm0.3)~\mathrm{MHz}$ and an effective dipole decay rate
$\gamma^\prime=2\pi\times(12.7\pm0.8)~\mathrm{MHz}$. Both values are consistent with the previous
measurement and the theoretical expectations. As in the previous measurement, the 894-nm
reference laser is used to compensate systematic drifts and acoustic vibrations. However, since
this compensation method relies on the temporal correlations of the drifts in both signals, and
thereby on their relative positions in the cavity scan, the compensation becomes less effective at
large detunings. This is reflected in the bigger spread and the larger error bars at larger
detunings, which renders this method slightly less precise than the absorption measurement to
evaluate the collective coupling rate.

\subsection{Vacuum Rabi splitting}
\label{sec:VaccumRabiSplitting}
\begin{figure}
  \centering
  \includegraphics[width=\columnwidth]{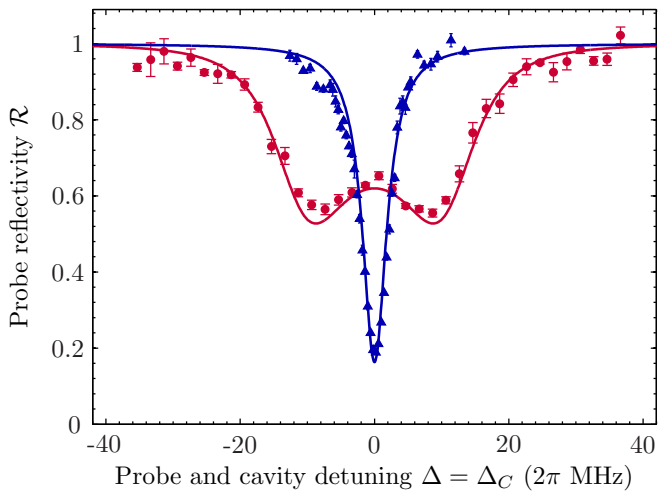}
  \caption{(color online). Probe reflectivity signal as a function of $\Delta=\Delta_{\mathrm{c}}$ for
    the empty cavity (blue triangles) and with a crystal with
    $N=520^{+24}_{-32}$ effectively interacting ions present in the cavity mode volume  (red
    circles). The solid lines are fits to the theory.
  \label{fig:rabisplitting}}
\end{figure}
A third complementary method to measure the collective coupling rate is based on locking the cavity
on atomic resonance, $\omega_{\mathrm{c}}=\omega_{\mathrm{at}}$. The response of the coupled atom-cavity system is then
probed as a function of probe detuning $\Delta$, which is then equal to the cavity detuning
$\Delta_{\mathrm{c}}$. The result of this measurement is shown on Fig.~\ref{fig:rabisplitting}. The blue triangles
are obtained with an empty cavity, while the red circles were taken with the same ion Coulomb
crystal as used in the previous experiments. Each data point is deduced from $2\times10^4$ experimental
sequences (see Fig.~\ref{fig:experimentalSequence}). The results are fitted using the theoretical
expectations of Eq.~\eqref{eq:reflectivity} and
  Eqs.~\eqref{eq:kappaPrimeTemperature} and \eqref{eq:delta_cprime_Temperature} (solid lines in Fig.~\ref{fig:rabisplitting}) and yield
$g_{\mathrm{N}}=2\pi\times(12.2\pm0.2)~\mathrm{MHz}$, a value that is in good agreement with the previous
measurements. To facilitate the convergence of the more complex fitting
function, the value of $\gamma^\prime$ in Eqs.~\eqref{eq:kappaPrimeTemperature}
and \eqref{eq:delta_cprime_Temperature} was set to the
one found in the previous absorption measurement.\\

From these three independent measurements of the collective coupling rate
$g_{\mathrm{N}}$ and using the effective number of ions $N=520^{+24}_{-32}$, one
deduces a single ion coupling rate of $g_{\mathrm{exp}}=2\pi\times
(0.53\pm0.02)~\mathrm{MHz}$, which
is in excellent agreement with the expected value of $g_{\mathrm{theory}}=2\pi\times(0.53\pm0.01)~\mathrm{MHz}$.

\subsection{Scaling with the number of interacting ions}
\label{sec:CouplingVsN}
\begin{figure}
  \centering
  \includegraphics[width=\columnwidth]{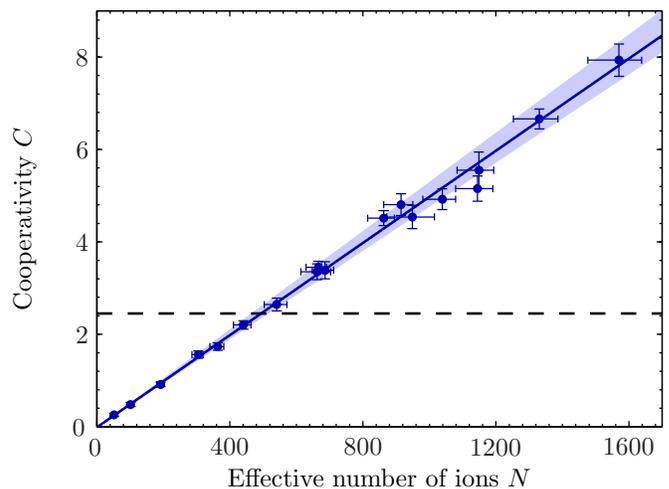}
  \caption{(Color online) Cooperativity as a function
      of the effective number of ions. The solid line is a linear fit to the
      data and yields a scaling parameter of $\frac C
      N=(5.1^{+0.4}_{-0.2})\times10^{-3}$. The shaded area indicates the
      confidence region for the upper and lower limit of $C$ determined by the
      (systematic) uncertainties on the effective number of ions, $N$. 
      The dashed line indicates the strong collective coupling limit
      $g_{\mathrm{N}}>(\kappa,\gamma)$. 
      \label{fig:CvsN}}
\end{figure}
To check further the agreement between the theoretical predictions and the experimental data, we
investigated the dependence of the collective coupling rate on the effective number of ions. An
attractive feature of ion Coulomb crystals is that the number of ions effectively interacting with
a single mode of the optical cavity can be precisely controlled by the trapping potentials. While
the density $\rho$ only depends on the amplitude of the RF voltage [see Eq.~\eqref{eq:ionDensity}],
the aspect ratio of the crystal depends on the relative trap depths of the axial and radial
confinement potentials, which can be independently controlled by the DC voltages on the endcap
electrodes. This allows for controlling the number of effectively interacting ions down to the few
ion-level.

By analogy with the case of a single two-level system interacting with a
  single field mode of an optical cavity, the cooperativity
  parameter $C$ is defined here as (half) the ratio of the square of the effective coupling rate $g_\mathrm{N}$ to the
cavity field decay rate $\kappa$ times the effective dipole decay rate
$\gamma^\prime$ (taking into account the effect of the motion of the ions):
$C=g_{\mathrm{N}}^2/2\kappa\gamma^\prime$. As can be seen from
Eq.~\eqref{eq:kappaPrimeTemperature}, this parameter can be experimentally
obtained by measuring for a probe field tuned to atomic resonance ($\Delta=0$)
the effective cavity field decay rate $\kappa^\prime(\Delta=0)=\kappa+\frac {g_{\mathrm{N}}^2}{\gamma^\prime}=\kappa~(1+2C)$.
In Fig.~\ref{fig:CvsN}, the dependence of the cooperativity parameter, $C$,
is plotted as a function of the effective number of ions interacting with the
TEM$_{00}$ mode, where the effective number of ions was changed by measuring for different aspect ratios and densities of several crystals.

The effective number of ions in each crystals was deduced by applying the method described in
Sec.~\ref{sec:effectivenumberofions}. The data points were obtained using
$\sigma^-$-circularly polarized probe light, hence probing the population in the $m_{\mathrm{J}}=+3/2$ and
$m_{\mathrm{J}}=+1/2$ substates, and shows the expected linear dependence on the effective number of ions.
From a linear fit (solid line) we deduce a scaling parameter $\frac C
      N=(5.1^{+0.4}_{-0.2})\times10^{-3}$. The limit where {\it
collective} strong coupling is achieved ($g_{\mathrm N}>\kappa,\gamma$) is
indicated by the black dashed line and is reached for $\approx 500$
interacting ions.\\
The largest coupling observed in these experiments was measured for a crystal with a length of
$\sim 3~\mathrm{mm}$ and a density of $\sim 6\times 10^8~\mathrm{cm^{-3}}$ and amounted to
$C=7.9\pm0.3$, corresponding to an effective number of ions of $N=1523^{+69}_{-93}$. This value exceeds
previously measured cooperativities with ions in optical cavities by roughly one order of
magnitude~\cite{Guthohrlein2001AsingleIon,Keller2004Continuous,Kreuter2004}. \\
\begin{figure}
  \includegraphics[width=\columnwidth]{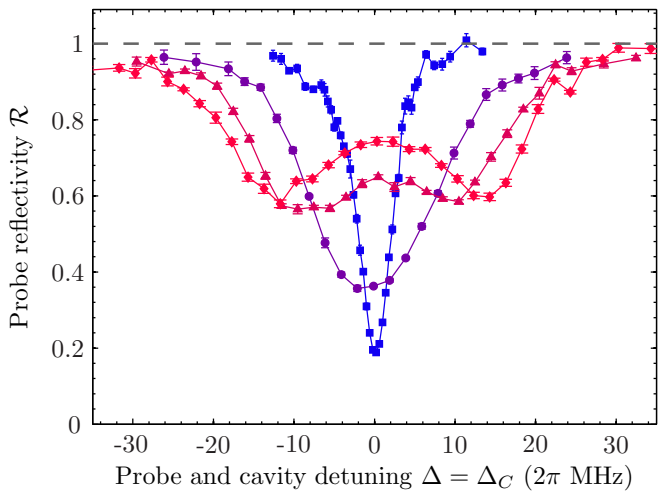}
  \caption{(color online). Vacuum Rabi splitting spectra ($\Delta=\Delta_{\mathrm{c}}$) obtained for increasing
    effective number of ions [0 (blue squares), 243 (lilac circles), 601 (dark
    red triangles), 914 (red diamonds)], the lines are presented to guide the eye.
  \label{fig:rabiFamily}}
\end{figure}

\begin{figure}
  \includegraphics[width=\columnwidth]{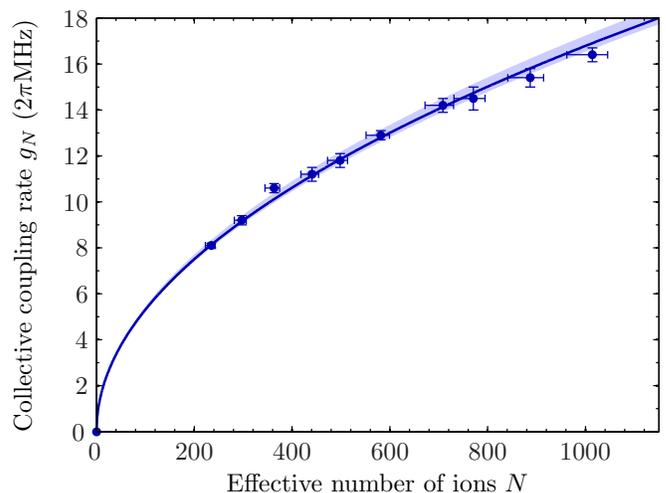}
  \caption{(color online). Collective coupling rate $g_{\mathrm N}$ versus effective number of ions
    $N$ deduced from reflectivity spectra, such as shown in Fig.~\ref{fig:rabiFamily}, obtained
    with crystals of different shape and density.
    The blue line is a fit to the data and gives a single ion coupling rate
    $g=2\pi\times(0.53\pm0.01)$~MHz. The shaded area indicates the lower and
    upper bound of the collective coupling rate $g_{\mathrm N}$ within the
    uncertainties of $N$. The horizontal errorbars are calculated according to Eq.~\eqref{eq:errorN}
  \label{fig:g_vs_N_exp}}
\end{figure}
Similarly, vacuum Rabi splitting spectra, such as the one presented in
Fig.~\ref{fig:rabisplitting}, were measured for several crystals and aspect ratios. The result of
such measurements is shown in Fig.~\ref{fig:rabiFamily}, showing clearly the increase in the
separation between the coupled crystal+cavity normal modes as the number of ions is increased. The
collective coupling rate $g_{\mathrm N}$, derived from fits to the theoretical
expression Eq. \eqref{eq:reflectivity}, is plotted for different effective number of ions in
Fig.~\ref{fig:g_vs_N_exp}. Taking the finite optical pumping efficiency into account and fitting
the curve with the expected square-root dependency, we deduce a single ion coupling rate of
$g=2\pi \times (0.53\pm0.01)~\mathrm{MHz}$, in good agreement with the previous measurements and
the theoretical expectation.

\section{Coherence time of collective Zeeman substate coherences}
\label{sec:coherencetime}
To evaluate the prospect for realizing coherent manipulations, we measured the
decay time of the collective coherences between the Zeeman substates of the
$3\mathrm{d}^2\mathrm{D}_{3/2}$ level. These coherences were established by
the Larmor precession of the magnetic spin induced by an additional $B$-field
transverse to the quantization axis. In presence 
of this orthogonal $B$-field, the population of the several substates 
undergo coherent oscillations, which are measured at different times in their
free evolution by directly probing the coherent coupling between the cavity field
and the ions. 
In order to be able to resolve the coherent
population oscillations in time using the previous technique (probing time $\sim
1~\mathrm{\mu s})$ the amplitude of the 
longitudinal $B$-field was lowered to obtain oscillation periods in the $\sim
10~\mathrm{\mu s}$ range, and the optical pumping preparation was modified as
to minimize the effect of the transverse $B$-field. 
The reduced $B$-field along the quantization axis could in principle make the sample more
sensitive to $B$-field fluctuations. Since these fluctuations might be one of the
factors eventually limiting the achievable coherence time, we expect the
coherence time measured by this method to be a lower bound as compared to the
previous configuration with a larger longitudinal $B$-field. 

\subsection{Experimental sequence and theoretical expectations}
\begin{figure*}%[htb]
  \centering
  \includegraphics[width=\textwidth]{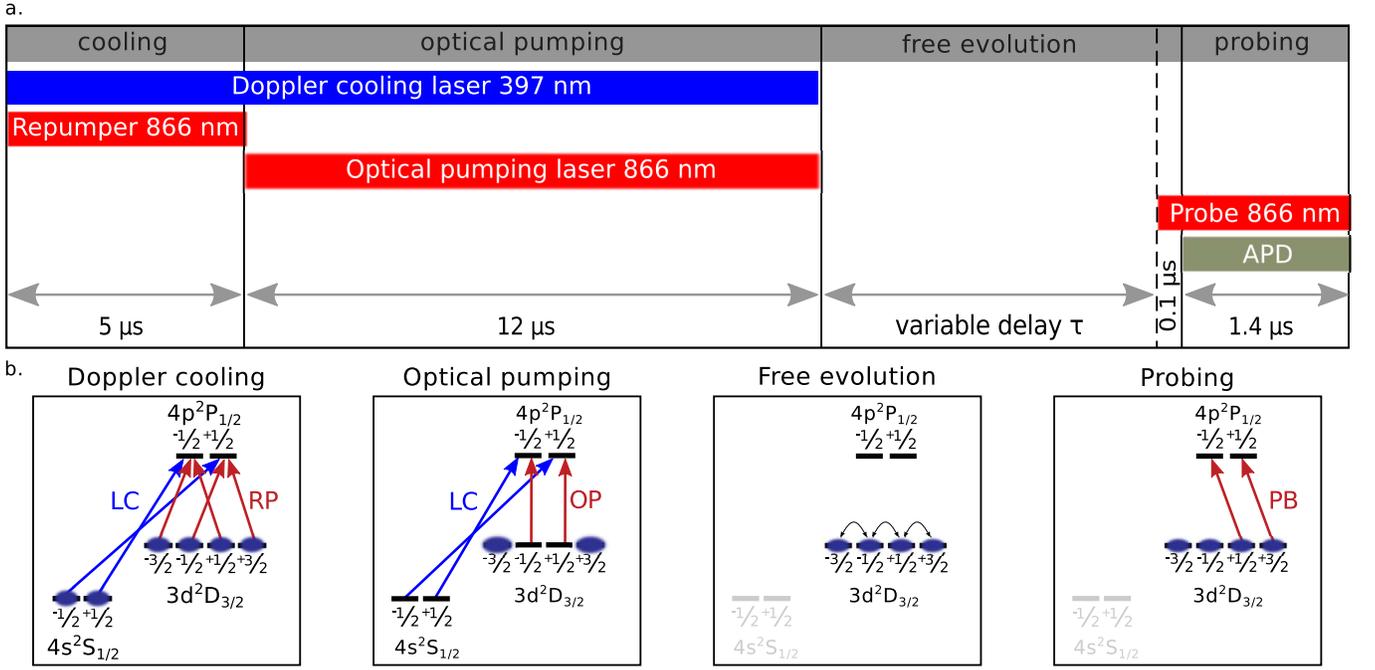}
  \caption{(Color online) (a) Experimental sequence used to measure the coherence
    time of collective Zeeman substate coherences in the
    $3\mathrm{d}^2\mathrm{D}_{3/2}$ level. (b) Energy levels of
    $^{40}\mathrm{Ca}^+$, including the relevant transitions and their
    polarization for the four phases of the experimental cycle. In the third
    phase, all lasers are turned off for a variable delay $\tau$, and the system
    evolves freely in presence of a transverse magnetic field component $B_{\mathrm{x}}$.
      \label{fig:coherenceTime_SequenceAndLevelScheme}}
\end{figure*}
The coherence time measurements required the experimental configuration and the measurement sequence
to be slightly modified as compared to the collective coupling rate
measurements described in Sec. \ref{sec:experimentalSequence}. The Larmor
precession is induced by an additional $B$-field component along the
transverse $x$ direction,
while the longitudinal magnetic field component $B_{\mathrm{z}}$ was lowered to optimize the contrast of the
coherent population oscillations. The optical pumping light propagates along
the $x$ axis and is $\pi$-polarized, hence transferring most of the atoms
symmetrically into the two outermost magnetic sub-states of the
$3\mathrm{d}^2\mathrm{D}_{3/2}$ level, $m_{\mathrm{J}}=\pm 3/2$. 

The experimental sequence used to measure the coherence time is shown in Fig.
\ref{fig:coherenceTime_SequenceAndLevelScheme}. The ions are Doppler-laser cooled during the first
$5~\mathrm{\mu s}$, followed by a $12~\mathrm{\mu s}$ optical pumping period. After the optical
pumping, all lasers are turned off for a time $\tau$, allowing for the free evolution of the
system. Finally, a weak $\sigma^-$-circularly polarized probe pulse is injected into the cavity,
addressing the ions in the $m_{\mathrm{J}}=+1/2$ and $m_{\mathrm{J}}=+3/2$ sub-states. The steady-state cavity reflection
is measured by collecting the reflected photons with the APD for $0.5~\mathrm{\mu s}$. The
additional delay time between optical pumping preparation and probing obviously lowers the
repetition rate of the sequence significantly, especially for long delay times, and the number of
data points for each sweep of the cavity will decrease. To compensate for
this, the data points at longer delays had to be averaged over more cavity scans, which substantially increased the acquisition
time and eventually limited these measurements to delays of around $\sim 120~\mathrm{\mu s}$.

Based on a simple four-level model the free Larmor precession-induced changes in the populations of
the Zeeman substates, $\ket{m_{\mathrm{J}}=\pm \nicefrac 1 2,~\pm \nicefrac 3 2}$, of the
$3\mathrm{d}^2\mathrm{D}_{3/2}$ level can be calculated. For a homogeneous $B$-field with
components $B_{\mathrm{x}}$ and $B_{\mathrm{z}}$, the Hamiltonian of the four-level system can be expressed in terms of
collective population operator, 
\begin{equation}
  \label{eq:larmorCollectivePopulationOperator}
  \hat \Pi_{{m_{\mathrm{J}}}}=\sum_{j=1}^{N_{\mathrm tot}}\ket
  {m_{\mathrm{J}}}^{(j)}\bra{m_{\mathrm{J}}}^{(j)},
\end{equation}
and collective spin operators
\begin{equation}
  \label{eq:larmorCollectiveSpinOperator}
  \hat \sigma_{{m_{\mathrm{J}}, m_{\mathrm{J}}^\prime}}=\sum_{j=1}^{N_{\mathrm
    tot}}\ket {m_{\mathrm{J}}}^{(j)}\bra{m_{\mathrm{J}}^\prime}^{(j)},\quad m_{\mathrm{J}}\ne m_{\mathrm{J}}^\prime.
\end{equation}
Here, $\ket{m_{\mathrm{J}}}^{(j)}$ and $\ket{m_{\mathrm{J}}^\prime}^{(j)}$ are the
state kets of the $j$th ion with magnetic quantum number $m_{\mathrm{J}}$ and
$m_{\mathrm{J}}^\prime$, respectively. The sum extends over the total
number of ions. In this notation, the Hamiltonian of the free evolution of a spin
$J=\nicefrac 3 2$ system can be written as
\begin{eqnarray}
  \nonumber
  H_{\mathrm{B}}&=&\hbar
  \omega_{\mathrm{z}}\sum_{m_{\mathrm{J}}}m_{\mathrm{J}}
  \hat\Pi_{{m_{\mathrm{J}}}} + \frac { \hbar
    \omega_{\mathrm{x}}} 2 \sum_{m_{\mathrm{J}}}\sum_{m_{\mathrm{J}}^\prime} \hat\sigma_{{m_{\mathrm{J}},m_{\mathrm{J}}^\prime}}\times\\
  &&   \label{eq:HamiltonianLarmor}
  \left[\sqrt{\frac {15}
      4 -m_{\mathrm{J}}(m_{\mathrm{J}}-1)} \delta_{{m_{\mathrm{J}},m_{\mathrm{J}}^\prime+1}} +\right. \\\nonumber
  && \left. \sqrt{\frac {15} 4 -m_{\mathrm{J}}(m_{\mathrm{J}}+1)} \delta_{{m_{\mathrm{J}},m_{\mathrm{J}}^\prime-1}} \right],
\end{eqnarray}
where the sums extend over the four Zeeman substates. Here, $\delta_{{m_{\mathrm{J}},m_{\mathrm{J}}^\prime+1}}$
is the Kronecker delta, and the Larmor frequencies $\omega_{\mathrm{z}}$ and
$\omega_{\mathrm{x}}$ corresponding to the $z$ and $x$ component of the magnetic field are given by the
product of the magnetic field amplitude by the gyromagnetic ratio $\gamma_{\mathrm{GM}}$:
\begin{equation}
  \label{eq:BField}
  \omega_{\mathrm{z}}=\gamma_{\mathrm{GM}} B_{\mathrm{z}},\quad \omega_{\mathrm{x}}=\gamma_{\mathrm{GM}} B_{\mathrm{x}}.
\end{equation}
For a $\sigma^-$-circularly polarized probe, the measured collective coupling to the cavity light
will depend on the
collective populations in the $m_{\mathrm{J}}=\nicefrac {+1} 2$ and $m_{\mathrm{J}}=\nicefrac {+3} 2$ substates. For a nonvanishing population
in the $m_{\mathrm{J}}=\nicefrac {+1} 2$ state, the measured effective cavity
decay rate, which was defined for a two-level system in Eq.~\eqref{eq:kappaPrime},
contains both contributions and is hence modified to
\begin{equation}
  \label{eq:kappaPrime_Larmor}
  \kappa^\prime(\tau)=\kappa+g_{\nicefrac 1 2}^2 N_{\nicefrac 1
    2}(\tau)\frac{\gamma}{\gamma^2+\Delta_{\nicefrac 1 2}^2}+ g_{\nicefrac 3 2}^2
  N_{\nicefrac 3 2}(\tau)\frac{\gamma}{\gamma^2+\Delta_{\nicefrac 3 2}^2},
\end{equation}
where $g_{m_{\mathrm{J}}}$, $N_{m_{\mathrm{J}}}$, and
$\Delta_{m_{\mathrm{J}}}=\omega_{m_{\mathrm{J}}}-\omega_{\mathrm{l}}$ denote
the single-ion coupling rate,  the effective number of ions and
the atomic detunings of the relevant Zeeman substates $m_{\mathrm{J}}=\nicefrac {+1}
2,~\nicefrac {+3} 2$, respectively, and $\omega_{m_J}$ is the frequency of the
$3\mathrm d^2 \mathrm D_{3/2},~m_{\mathrm{J}}\leftrightarrow
4\mathrm p^2\mathrm P_{1/2},m_{\mathrm{J}}-1$ transition. 

Due to the induced Larmor precession, the effective number of ions in the
individual Zeeman substates will be time-dependent. For a system initially
prepared in a superposition state $\psi_0$, the population in a particular
Zeeman substate at a certain time
$\tau$ can be calculated from the projection of the time evolved state,
$\psi(\tau)=U(\tau) \psi_0$, onto this state. Here, $U(\tau)=exp(-\nicefrac i \hbar H_{\mathrm{B}}
\tau)$ denotes the time evolution operator.
Straightforward but lengthy calculations show that the populations in the
$\nicefrac {+1} 2$ and $\nicefrac {+3} 2$ Zeeman substates after a time $\tau$ are of the form
$A\cos(\omega_{\mathrm{L}}\tau)+B\cos(2\omega_{\mathrm{L}}\tau)+C$, where $A$, $B$, and $C$ are
constants depending on the efficiency of the optical pumping (i.e., the initial
populations and coherences in the different Zeeman sublevels) and the magnetic
field amplitudes $Bz$ and $B_x$ (via $\omega_{\mathrm{x}}$ and $\omega_{\mathrm{z}}$). One thus
obtains $N_{1/2}(\tau)$ and $N_{3/2}(\tau)$ using Eq. (8). It follows from
Eq. \eqref{eq:HamiltonianLarmor} and $\kappa'(\tau)=\kappa(1+2C(\tau))$ that the measured
cooperativity at time $\tau$ can be put under the form
\begin{equation} 
  \label{eq:cooperativity_Larmor}
  C(\tau)=a \cos(\omega_{\mathrm{L}}\tau)+b\cos(2\omega_{\mathrm{L}}\tau)+c,
\end{equation}
where the Larmor frequency
\begin{equation}
  \label{eq:LarmorFrequency}
 \omega_{\mathrm{L}}=\sqrt{\omega_{\mathrm{z}}^2+\omega_{\mathrm{x}}^2}
\end{equation}
was defined. The parameters $a$, $b$, $c$ are constants depending on the efficiency of the optical
pumping preparation, and the magnetic field amplitudes $B_{\mathrm{z}}$ and $B_{\mathrm{x}}$.

\subsection{Experimental results}

\begin{figure}[htb]
  \centering
  \includegraphics[width=\columnwidth]{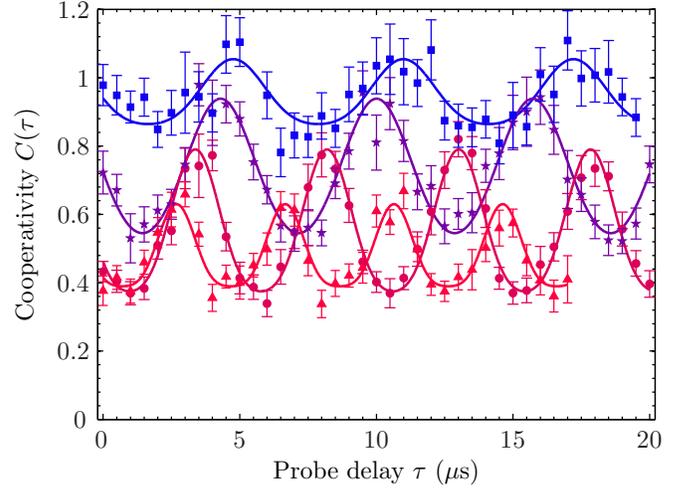}
  \caption{(Color online) Calibration of the Larmor
      frequency for different currents of the $B_{\mathrm{x}}$ coils. Shown is the
      cooperativity as a function of delay time $\tau$ for different
      transverse $B$-fields: $I_{\mathrm{x}}=10~\mathrm{mA}$
      (blue squares), $I_{\mathrm{x}}=16~\mathrm{mA}$ (lilac stars), $I_{\mathrm{x}}=26~\mathrm{mA}$
      (dark red circles), and
      $I_{\mathrm{x}}=36~\mathrm{mA}$ (red triangles). The solid lines are fits according to
      Eq.~\eqref{eq:cooperativity_Larmor}.
      \label{fig:BFieldCalibration}}
\end{figure}
\begin{figure}[htb]
  \centering
  \includegraphics[width=\columnwidth]{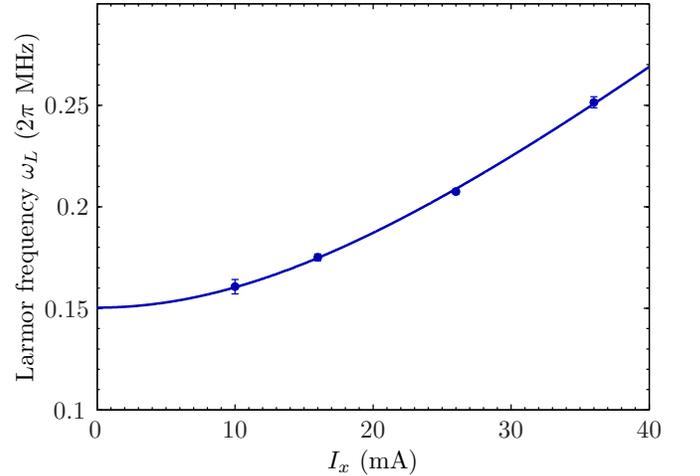}
  \caption{(Color online) Larmor frequency as a function
      of current through the $B_{\mathrm{x}}$ coils. The solid line is a fit of the
      form $\omega_{\mathrm{L}}=\sqrt{\omega_{\mathrm{z}}^2+a^2 I_{\mathrm{x}}^2}$ and we deduce
      $\omega_{\mathrm{z}}=2\pi\times (0.150\pm 0.002)~\mathrm{MHz}$ and
      $\omega_{\mathrm{x}}=2\pi\times(5.5\pm0.1)~\frac{\mathrm{kHz}}{\mathrm{mA}}\times I_{\mathrm{x}}$.
      \label{fig:BFieldCalibration2}}
\end{figure}

The amplitudes of the magnetic fields, $B_{\mathrm{x}}$ and $B_{\mathrm{z}}$,
at the position of the ion crystal were calibrated
by measuring the dependence of the Larmor frequency $\omega_{\mathrm{L}}$ with the intensity of the current
used to drive the transverse magnetic field coils [see Eqs. \eqref{eq:BField} and
\eqref{eq:LarmorFrequency}]. The obtained coupling as a function of $\tau$ is shown for different
currents $I_{\mathrm{x}}$ on Fig.~\ref{fig:BFieldCalibration}. The curves are fitted
according to Eq.~\eqref{eq:cooperativity_Larmor}, yielding the individual
Larmor frequencies. These frequencies are shown as a 
function of the current through the $B_{\mathrm{x}}$ coils in Fig.~\ref{fig:BFieldCalibration2}. Using the
gyromagnetic ratio $\gamma_{\mathrm{GM}}=\frac {\mu_{\mathrm{B}} \mathfrak{g}_{3/2}} {\hbar}$ ($\mu_{\mathrm{B}}$ is the Bohr
magneton, $\mathfrak{g}_{3/2}$ the Land\'e factor of the $3\mathrm{d}^2\mathrm{D}_{3/2}$ level), we deduce the
magnetic fields along the two axis $B_{\mathrm{z}}=(0.134\pm0.002)~\mathrm{G}$ and
$B_{\mathrm{x}}=(4.91\pm0.09)~\frac{\mathrm{G}}{\mathrm{A}}\times I_{\mathrm{x}}$.\\

\begin{figure}[htb]
  \centering
    \centering
    \includegraphics[width=\columnwidth]{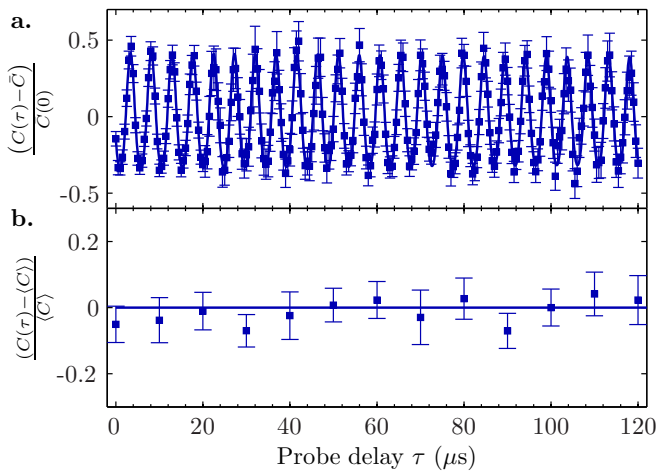}
    \caption{(Color online) (a) Normalized cooperativity
      paramter as a function of delay $\tau$. Due to the presence of a
      non-zero $B$-field component orthogonal to the quantization axis 
      ($B_{\mathrm{z}}=B_{\mathrm{x}}=0.15~\mathrm{G}$), coherent Larmor precessions are observed. Long term
      drifts are compensated by normalizing to the mean of one oscillation
      period. The solid line corresponds to a fit, assuming an exponential
      decay and yields a coherence time of
      $\tau_{\mathrm{e}}=1.7^{100}_{-0.8}~\mathrm{ms}$.
      (b) Cooperativity as a function of delay with a $B$-field
      present only along the quantization axis
      ($B_{\mathrm{x}}=B_{\mathrm{y}}=0,~B_{\mathrm{z}}=0.15~\mathrm{G}$). The data 
      points are normalized to the mean cooperativity of $\langle C
      \rangle=1.43\pm 0.02$.
      \label{fig:coherenceConstantCoupling}
      \label{fig:coherenceTime}}
\end{figure}
To achieve a large contrast, the measurement was carried out with moderate $B$-field values
$B_{\mathrm{x}}=B_{\mathrm{z}}=0.15~\mathrm{G}$ and the variation of the cooperativity was measured for $120~\mathrm{\mu
  s}$. To compensate for slow drifts during the measurement, each
data point was normalized to the mean cooperativity, $\bar C$, averaged over one oscillation
period. The normalized cooperativity is shown in Fig. \ref{fig:coherenceTime}(a), together with a
fit of the form of \eqref{eq:cooperativity_Larmor}, where decoherence processes are taken into
account by multiplying the oscillating terms with an exponential decay term
$\exp(-\nicefrac {\tau} {\tau_{\mathrm{e}}})$,
which would be expected, e.g., for a homogeneous broadening of the energy levels. From this fit, we
deduce a coherence time of $\tau_{\mathrm{e}}=1.7^{100}_{-0.8}~\mathrm{ms}$. This value is comparable to
previously measured coherence times for single ions in linear Paul trap in equivalent magnetic
field sensitive states \cite{Schmidt-Kaler2003} and might be further improved by an active control
of stray magnetic fields or state configurations that are less magnetic field sensitive. For
inhomogeneous broadening, due to magnetic field gradient over the crystal, the decoherence process
would be better described by a Gaussian decay~\cite{Chaneliere2005}. Fitting the data assuming a Gaussian decay
$\exp(- \nicefrac {\tau^2}{\tau_{\mathrm{g}}^2})$ in Eq.~\eqref{eq:cooperativity_Larmor} yields a coherence time of
$\tau_{\mathrm{g}}=0.5_{-0.2}^{+0.6}~\mathrm{ms}$. Due to the limitation of our measurement to time delays of
$\tau \lesssim 120~\mathrm{\mu s}$, it is at present not possible to distinguish between the two decay
mechanisms.

For comparison, the cooperativity as a function of probe delay, $C(\tau)$, was measured with only
the bias field along the quantization axis present ($B_{\mathrm{x}}=0,~B_{\mathrm{z}}=0.15\mathrm{G}$), as shown in Fig.
\ref{fig:coherenceConstantCoupling} b. Here, the values are normalized to the mean cooperativity
averaged over all points $\langle C \rangle$. Within the error bars, the deduced cooperativities
agree with a constant value of $\langle C \rangle=1.43\pm 0.02$ (solid line).

\section{Conclusion}\label{sec:conclusion}

To conclude, we have presented a detailed theoretical and experimental
analysis of the experiments of~\cite{Herskind2009Realization}, which
demonstrated the possibility of using large ion Coulomb crystals positioned in
a moderately high-finesse optical cavity to enter the collective  
strong-coupling regime of CQED.
The excellent agreement between the experimental results including those of
Ref.~\cite{Herskind2009Realization}
 and the theoretical predictions, makes ion Coulomb crystals promising candidates for the
realization of quantum information processing devices such as quantum memories and
repeaters~\cite{Duan2001,Kimble2008}. Using, for instance, cavity EIT-based
protocols~\cite{Fleischhauer2000,Lukin2000,Dantan2008c,Gorshkov2007,Albert2011Cavity}, the obtained coupling
strengths and coherence times could open up for the realization of both high-efficiency
\textit{and} long life-time quantum memories~\cite{Lvovsky2009Optical}. Moreover, the nice
properties of ion Coulomb crystals also allow for the manipulation of complex multimode photonic
information~\cite{Lvovsky2009Optical} by exploiting the crystal spatial~\cite{Dantan2009Large} or
motional~\cite{Dantan2010Non-invasive} degrees of freedom. Ion Coulomb crystals in optical cavities
have also great potential for the investigation of cavity optomechanical
phenomena~\cite{Kippenberg2008} and the observation of novel phase
transitions~\cite{Garcia-Mata2007Frenkel-Kontorova,Retzker2008Double,Fishman2008Structural,Harkonen2009Dicke,Baumann2010Dicke}
with cold, solid-like objects.

We acknowledge financial support from the Carlsberg Foundation,
the Danish Natural Science Research Council through the ESF EuroQUAM
project CMMC, and the EU commission through the FP7 ITN project CCQED and STREP project PICC.\\


\begin{thebibliography}{10}

\bibitem{Berman1994}
P.~R. Berman, editor,
\newblock {\em {Cavity Quantum Electrodynamics}} (Academic Press, London, 1994).

\bibitem{Haroche2006}
S.~Haroche and J.-M. Raimond,
\newblock {\em {Exploring the Quantum: Atoms, Cavities, and Photons}} (Oxford
  University Press, Oxford, 2006).

\bibitem{Rempe1994Cavity}
G.~Rempe, R.~J. Thompson, and H.~J. Kimble,
\newblock Phys. Scr. {\bf 1994}, 67 (1994).

\bibitem{Brune1996}
M.~Brune, A.~Maali, J.~Dreyer, E.~Hagley, J.~M.~Raimond, and S.~Haroche,
\newblock Phys. Rev. Lett. {\bf 76}, 1800 (1996).

\bibitem{Thompson1992Observation}
R.~J. Thompson, G.~Rempe, and H.~J. Kimble,
\newblock Phys. Rev. Lett. {\bf 68}, 1132 (1992).

\bibitem{Badolato2005Deterministic}
A.~Badolato, K.~Hennessy, M.~Atat\"{u}re, J.~Dreiser, E.~Hu, P.~M.~Petroff, and A.~Imamo\u{g}lu,
\newblock Science {\bf 308}, 1158 (2005).

\bibitem{khitrova2006}
G.~Khitrova, H.~M. Gibbs, M.~Kira, S.~W. Koch, and A.~Scherer,
\newblock Nature Phys. {\bf 2}, 81 (2006).

\bibitem{wallraff2004}
A.~Wallraff, D.~I.~Schuster, A.~Blais, L.~Frunzio, R.-S.~Huang, J.~Majer, S.~Kumar, S.~M.~Girvin, and R.~J.~Schoelkopf,
\newblock Nature (London) {\bf 431}, 162 (2004).

\bibitem{Chiorescu2004Coherent}
I.~Chiorescu, P.~Bertet, K.~Semba, Y.~Nakamura, C.~J.~P.~M.~Harmans, and J.~E.~Mooij,
\newblock Nature (London) {\bf 431}, 159 (2004).

\bibitem{Hood1998}
C.~J. Hood, M.~S. Chapman, T.~W. Lynn, and H.~J. Kimble,
\newblock Phys. Rev. Lett. {\bf 80}, 4157 (1998).

\bibitem{Maunz2005}
P.~Maunz, T.~Puppe, I.~Schuster, N.~Syassen, P.~W.~H.~Pinkse, and G.~Rempe,
\newblock Phys. Rev. Lett. {\bf 94}, 033002 (2005).

\bibitem{Harlander2010Trapped-ion}
M.~Harlander, M.~Brownnutt, W.~H\"{a}nsel, and R.~Blatt,
\newblock New J. Phys {\bf 12}, 093035+ (2010).

\bibitem{Herskind2011AMicrofabricated}
P.~F.~Herskind, S.~X.~Wang, M.~Shi, Y.~Ge, M.~Cetina, and I.~L.~Chuang,
\newblock Opt. Lett. {\bf 36}, 3045 (2011).

\bibitem{Guthohrlein2001AsingleIon}
G.~R. Guthohrlein, M.~Keller, K.~Hayasaka, W.~Lange, and H.~Walther,
\newblock Nature {\bf 414}, 49 (2001).

\bibitem{Kreuter2004}
A.~Kreuter, C.~Becher, G.~P.~T.~Lancaster, A.~B.~Mundt, C.~Russo, H.~Haffner,
C.~Roos, J.~Eschner, F.~Schmidt-Kaler, and R.~Blatt,
\newblock Phys. Rev. Lett. {\bf 92}, 203002 (2004).

\bibitem{Keller2004Continuous}
M.~Keller, B.~Lange, K.~Hayasaka, W.~Lange, and H.~Walther,
\newblock Nature (London) {\bf 431}, 1075 (2004).

\bibitem{Barros2009Deterministic}
H.~G.~Barros, A.~Stute, T.~E.~Northup, C.~Russo, P.~O.~Schmidt, and R.~Blatt,
\newblock New J. Phys. {\bf 11}, 103004+ (2009).

\bibitem{Leibrandt2009Cavity}
D.~R. Leibrandt, J.~Labaziewicz, V.~Vuleti\'{c}, and I.~L. Chuang,
\newblock Phys. Rev. Lett. {\bf 103}, 103001+ (2009).

\bibitem{Dubin2010Quantum}
F.~Dubin, C.~Russo, H.~G.~Barros, A.~Stute, C.~Becher, P.~O.~Schmidt, and R.~Blatt,
\newblock Nature Phys. {\bf 6}, 350 (2010).

\bibitem{Kaluzny1983}
Y.~Kaluzny, P.~Goy, M.~Gross, J.~M. Raimond, and S.~Haroche,
\newblock Phys. Rev. Lett. {\bf 51}, 1175 (1983).

\bibitem{Lambrecht1996}
A.~Lambrecht, T.~Coudreau, A.~M. Steinberg, and E.~Giacobino,
\newblock Europhys. Lett. {\bf 36}, 93 (1996).

\bibitem{Nagorny2003Collective}
B.~Nagorny, T.~Els\"{a}sser, and A.~Hemmerich,
\newblock Phys. Rev. Lett. {\bf 91}, 153003+ (2003).

\bibitem{Chan2003Observation}
H.~W. Chan, A.~T. Black, and V.~Vuleti\'{c},
\newblock Phys. Rev. Lett. {\bf 90}, 063003+ (2003).

\bibitem{Kruse2004Observation}
D.~Kruse, C.~von Cube, C.~Zimmermann, and P.~W. Courteille,
\newblock Phys. Rev. Lett. {\bf 91}, 183601+ (2003).

\bibitem{Chen2011Conditional}
Z.~Chen, J.~G. Bohnet, S.~R. Sankar, J.~Dai, and J.~K. Thompson,
\newblock Phys. Rev. Lett. {\bf 106}, 133601+ (2011).

\bibitem{Brennecke2007Cavity}
F.~Brennecke, T.~Donner, S.~Ritter, T.~Bourdel, M.~K\"{o}hl, and T.~Esslinger,
\newblock Nature (London) {\bf 450}, 268 (2007).

\bibitem{Colombe2007Strong}
Y.~Colombe, T.~Steinmetz, G.~Dubois, F.~Linke, D.~Hunger, and J.~Reichel,
\newblock Nature (London) {\bf 450}, 272 (2007).

\bibitem{Herskind2009Realization}
P.~F. Herskind, A.~Dantan, J.~P. Marler, M.~Albert, and M.~Drewsen,
\newblock Nature Phys. {\bf 5}, 494 (2009).

\bibitem{Kimble2008}
H.~J. Kimble,
\newblock Nature (London) {\bf 453}, 1023 (2008).

\bibitem{Lambrecht1995Optical}
A.~Lambrecht, E.~Giacobino, and J.~M. Courty,
\newblock Opt. Commun. {\bf 115}, 199 (1995).

\bibitem{Joshi2003Optical}
A.~Joshi and M.~Xiao,
\newblock Phys. Rev. Lett. {\bf 91}, 143904+ (2003).

\bibitem{Grangier1991Observation}
P.~Grangier, J.~F. Roch, and G.~Roger,
\newblock Phys. Rev. Lett. {\bf 66}, 1418 (1991).

\bibitem{Roch1997Quantum}
J.~F. Roch, , K.~Vigneron, P.~Grelu, A.~Sinatra, J.~P.~Poizat, and P.~Grangier,
\newblock Phys. Rev. Lett. {\bf 78}, 634 (1997).

\bibitem{Mielke1998Nonclassical}
S.~L. Mielke, G.~T. Foster, and L.~A. Orozco,
\newblock Phys. Rev. Lett. {\bf 80}, 3948 (1998).

\bibitem{Black2005On-Demand}
A.~T. Black, J.~K. Thompson, and V.~Vuleti\'{c},
\newblock Phys. Rev. Lett. {\bf 95}, 133601+ (2005).

\bibitem{Thompson2006AHigh-Brightness}
J.~K. Thompson, J.~Simon, H.~Loh, and V.~Vuletic,
\newblock Science {\bf 313}, 74 (2006).

\bibitem{Simon2007}
J.~Simon, H.~Tanji, S.~Ghosh, and V.~Vuletic,
\newblock Nature Phys. {\bf 3}, 765 (2007).

\bibitem{Tanji2009Heralded}
H.~Tanji, S.~Ghosh, J.~Simon, B.~Bloom, and V.~Vuleti\'{c},
\newblock Phys. Rev. Lett. {\bf 103}, 043601+ (2009).

\bibitem{Josse2003Polarization}
V.~Josse,  A.~Dantan, L.~Vernac, A.~Bramati, M.~Pinard, and E.~Giacobinoe,
\newblock Phys. Rev. Lett. {\bf 91}, 103601+ (2003).

\bibitem{Josse2004Continuous}
V.~Josse, A.~Dantan, A.~Bramati, M.~Pinard, and E.~Giacobino,
\newblock Phys. Rev. Lett. {\bf 92}, 123601+ (2004).

\bibitem{Leroux2010Implementation}
I.~D. Leroux, M.~H. Schleier-Smith, and V.~Vuleti\'{c},
\newblock Phys. Rev. Lett. {\bf 104}, 073602+ (2010).

\bibitem{Klinner2006Normal}
J.~Klinner, M.~Lindholdt, B.~Nagorny, and A.~Hemmerich,
\newblock Phys. Rev. Lett. {\bf 96}, 023002+ (2006).

\bibitem{Slama2007Superradiant}
S.~Slama, S.~Bux, G.~Krenz, C.~Zimmermann, and P.~W. Courteille,
\newblock Phys. Rev. Lett. {\bf 98}, 053603+ (2007).

\bibitem{Murch2008Observation}
K.~W. Murch, K.~L. Moore, S.~Gupta, and D.~M. Stamper-Kurn,
\newblock Nature Phys. {\bf 4}, 561 (2008).

\bibitem{Brennecke2008}
F.~Brennecke, S.~Ritter, T.~Donner, and T.~Esslinger,
\newblock Science {\bf 322}, 235 (2008).

\bibitem{Black2003Observation}
A.~T. Black, H.~W. Chan, and V.~Vuleti\'{c},
\newblock Phys. Rev. Lett. {\bf 91}, 203001+ (2003).

\bibitem{Baumann2010Dicke}
K.~Baumann, C.~Guerlin, F.~Brennecke, and T.~Esslinger,
\newblock Nature (London) {\bf 464}, 1301 (2010).

\bibitem{Lange2009CavityQED}
W.~Lange,
\newblock Nature Phys. {\bf 5}, 455 (2009).

\bibitem{Lukin2000}
M.~D. Lukin, S.~F. Yelin, and M.~Fleischhauer,
\newblock Phys. Rev. Lett. {\bf 84}, 4232 (2000).

\bibitem{Duan2001}
L.~M. Duan, M.~D. Lukin, J.~I. Cirac, and P.~Zoller,
\newblock Nature (London) {\bf 414}, 413 (2001).

\bibitem{Leibfried2003Quantum}
D.~Leibfried, R.~Blatt, C.~Monroe, and D.~Wineland,
\newblock Rev. Mod. Phys {\bf 75}, 281 (2003).

\bibitem{Blatt2008Entangled}
R.~Blatt and D.~Wineland,
\newblock Nature (London) {\bf 453}, 1008 (2008).

\bibitem{Drewsen1998}
M.~Drewsen, C.~Brodersen, L.~Hornek\ae{}r, J.~S. Hangst, and J.~P. Schiffer,
\newblock Phys. Rev. Lett. {\bf 81}, 2878 (1998).

\bibitem{Hornekaer2002Formation}
L.~Hornek{\ae}r and M.~Drewsen,
\newblock Phys. Rev. A {\bf 66}, 013412+ (2002).

\bibitem{Hornekaer2001}
L.~Hornek\ae{}r, N.~Kj\ae{}rgaard, A.~M. Thommesen, and M.~Drewsen,
\newblock Phys. Rev. Lett. {\bf 86}, 1994 (2001).

\bibitem{Dantan2009Large}
A.~Dantan, M.~Albert, J.~P. Marler, P.~F. Herskind, and M.~Drewsen,
\newblock Phys. Rev. A {\bf 80}, 041802+ (2009).

\bibitem{Lvovsky2009Optical}
A.~I. Lvovsky, B.~C. Sanders, and W.~Tittel,
\newblock Nature Photon. {\bf 3}, 706 (2009).

\bibitem{Vasilyev2008Quantum}
D.~V. Vasilyev, I.~V. Sokolov, and E.~S. Polzik,
\newblock Phys. Rev. A {\bf 77}, 020302+ (2008).

\bibitem{Tordrup2008Holographic}
K.~Tordrup, A.~Negretti, and K.~M{\o}{}lmer,
\newblock Phys. Rev. Lett. {\bf 101}, 040501+ (2008).

\bibitem{Wesenberg2011Dynamics}
J.~H. Wesenberg, Z.~Kurucz, and K.~M{\o}{}lmer,
\newblock Phys. Rev. A {\bf 83}, 023826+ (2011).

\bibitem{Dubin1991Theory}
D.~H.~E. Dubin,
\newblock Phys. Rev. Lett. {\bf 66}, 2076 (1991).

\bibitem{Dubin1996Normal}
D.~H.~E. Dubin and J.~P. Schiffer,
\newblock Phys. Rev. E {\bf 53}, 5249 (1996).

\bibitem{Dantan2010Non-invasive}
A.~Dantan, J.~P. Marler, M.~Albert, D.~Gu\'{e}not, and M.~Drewsen,
\newblock Phys. Rev. Lett. {\bf 105}, 103001+ (2010).

\bibitem{Kippenberg2008}
T.~J. Kippenberg and K.~J. Vahala,
\newblock Science {\bf 321}, 1172 (2008).

\bibitem{Garcia-Mata2007Frenkel-Kontorova}
I.~Garc\'{\i}a-Mata, O.~V. Zhirov, and D.~L. Shepelyansky,
\newblock Eur. Phys. J. D {\bf 41}, 325 (2007).

\bibitem{Retzker2008Double}
A.~Retzker, R.~C. Thompson, D.~M. Segal, and M.~B. Plenio,
\newblock Phys. Rev. Lett. {\bf 101}, 260504+ (2008).

\bibitem{Fishman2008Structural}
S.~Fishman, G.~De~Chiara, T.~Calarco, and G.~Morigi,
\newblock Phys. Rev. B {\bf 77}, 064111+ (2008).

\bibitem{Harkonen2009Dicke}
K.~H\"{a}rk\"{o}nen, F.~Plastina, and S.~Maniscalco,
\newblock Phys. Rev. A {\bf 80}, 033841+ (2009).

\bibitem{Breuer2007TheTheory}
H.-P. Breuer and F.~Petruccione,
\newblock {\em The Theory of Open Quantum Systems} (Oxford University Press, Oxford, 2007).

\bibitem{Kogelnik1966Laser}
H.~Kogelnik and T.~Li,
\newblock Appl. Opt. {\bf 5}, 1550 (1966).

\bibitem{Scully1997}
M.~Scully and M.~Zubairy,
\newblock {\em {Quantum Optics}} (Cambridge University Press, Cambridge, 1997).

\bibitem{Tavis1968Exact}
M.~Tavis and F.~W. Cummings,
\newblock Phys. Rev. {\bf 170}, 379 (1968).

\bibitem{Raizen1989Normal-mode}
M.~G. Raizen, R.~J. Thompson, R.~J. Brecha, H.~J. Kimble, and H.~J. Carmichael,
\newblock Phys. Rev. Lett. {\bf 63}, 240 (1989).

\bibitem{Herskind2008Loading}
P.~Herskind, A.~Dantan, M.~B.~Langkilde-Lauesen, A.~Mortensen, J.~L.~S\o rensen, and M.~Drewsen,
\newblock Appl. Phys. B {\bf 93}, 373 (2008).

\bibitem{Kjaergaard2000Isotope}
N.~Kj{\ae}rgaard, L.~Hornek{\ae}r, A.~M. Thommesen, Z.~Videsen, and M.~Drewsen,
\newblock Appl. Phys. B {\bf 71}, 207 (2000).

\bibitem{Mortensen2004Isotope}
A.~Mortensen, J.~J.~T. Lindballe, I.~S.~Jensen, P.~Staanum, D.~Voigt, and M.~Drewsen,
\newblock Phys. Rev. A {\bf 69}, 042502+ (2004).

\bibitem{Drever1983Laser}
R.~W.~P.~Drever, J.~L.~Hall, F.~V.~Kowalski, J.~Hough, G.~M.~Ford, A.~J.~Munley, and H.~Ward,
\newblock Appl. Phys. B {\bf 31}, 97 (1983).

\bibitem{phdPeterHerskind}
P.~Herskind,
\newblock {\em {Cavity Quantum Electrodynamics with Ion Coulomb Crystals}},
\newblock PhD thesis, The University of Aarhus, 2008.

\bibitem{Turner1987Collective}
L.~Turner,
\newblock Phys. Fluids {\bf 30}, 3196 (1987).

\bibitem{Herskind2009Positioning}
P.~F. Herskind, A.~Dantan, M.~Albert, J.~P. Marler, and M.~Drewsen,
\newblock J. of Phys. B {\bf 42}, 154008+ (2009).

\bibitem{Schmidt-Kaler2003}
F.~Schmidt-Kaler, S.~Gulde, M.~Riebe, T.~Deuschle, A.~Kreuter, G.~Lancaster, C.~Becher, J.~Eschner, H.~H\"{a}ffner, and R.~Blatt,
\newblock J. Phys. B {\bf 36}, 623 (2003).

\bibitem{Chaneliere2005}
T.~Chaneli\`{e}re, D.~N.~Matsukevich, S.~D.~Jenkins, S.-Y.~Lan, T.~A.~B.~Kennedy, and A.~Kuzmich,
\newblock Nature (London) {\bf 438}, 833 (2005).

\bibitem{Fleischhauer2000}
M.~Fleischhauer, S.~F. Yelin, and M.~D. Lukin,
\newblock Opt. Commun. {\bf 179}, 395 (2000).

\bibitem{Dantan2008c}
A.~Dantan and M.~Pinard,
\newblock Phys. Rev. A {\bf 69}, 043810 (2004).

\bibitem{Gorshkov2007}
A.~V. Gorshkov, A.~Andr\'{e}, M.~D. Lukin, and A.~S. S\o{}rensen,
\newblock Phys. Rev. A {\bf 76}, 033804 (2007).

\bibitem{Albert2011Cavity}
M.~Albert, A.~Dantan, and M.~Drewsen,
\newblock Nature Photon. {\bf 5}, 633 (2011).

\end{thebibliography}
\end{document}